\documentclass[letterpaper,pop,showpacs,preprint,superscriptaddress]{revtex4-1}
\usepackage{lmodern}

\usepackage[T1]{fontenc}
\usepackage[latin9]{inputenc}
\usepackage{amsmath}
\usepackage{amssymb}
\usepackage{graphicx}

\makeatletter

\usepackage{amsthm}\usepackage{latexsym}\usepackage{bm}\usepackage{amsfonts}

\makeatother

\begin{document}

\title{One-dimensional kinetic description of nonlinear traveling-pulse
(soliton) and traveling-wave disturbances in long coasting charged
particle beams}

\author{Ronald C. Davidson}

\affiliation{Plasma Physics Laboratory, Princeton University, Princeton, New Jersey
08543, USA}

\author{Hong Qin}

\affiliation{Plasma Physics Laboratory, Princeton University, Princeton, New Jersey
08543, USA}

\affiliation{School of Nuclear Science and Technology and Department of Modern
Physics, University of Science and Technology of China, Hefei, Anhui
230026, China}
\begin{abstract}
This paper makes use of a one-dimensional kinetic model to investigate
the nonlinear longitudinal dynamics of a long coasting beam propagating
through a perfectly conducting circular pipe with radius $r_{w}.$
The average axial electric field is expressed as $\langle E_{z}\rangle=-(\partial/\partial z)\langle\phi\rangle=-e_{b}g_{0}\partial\lambda_{b}/\partial z-e_{b}g_{2}r_{w}^{2}\partial^{3}\lambda_{b}/\partial z^{3}$,
where $g_{0}$ and $g_{2}$ are constant geometric factors, $\lambda_{b}(z,t)=\int dp_{z}F_{b}(z,p_{z},t)$
is the line density of beam particles, and $F_{b}(z,p_{z},t)$ satisfies
the 1D Vlasov equation. Detailed nonlinear properties of traveling-wave
and traveling-pulse (solitons) solutions with time-stationary waveform
are examined for a wide range of system parameters extending from
moderate-amplitudes to large-amplitude modulations of the beam charge
density. Two classes of solutions for the beam distribution function
are considered, corresponding to: (a) the nonlinear waterbag distribution,
where $F_{b}=const.$ in a bounded region of $p_{z}$-space; and (b)
nonlinear Bernstein-Green-Kruskal (BGK)-like solutions, allowing for
both trapped and untrapped particle distributions to interact with
the self-generated electric field $\langle E_{z}\rangle$.
\end{abstract}


\maketitle

\section{Introduction\label{sec:Introduction}}

High-energy accelerators and transport systems \cite{1davidson2001physics,2reiser2008theory,3chao1993physics,4edwards55163introduction,Lawson88,Wangler98}
have a wide variety of applications ranging from basic research in
high energy and nuclear physics, to applications such as spallation
neutron sources, medical physics, and heavy ion fusion. As a consequence,
it is increasingly important to develop an improved understanding
of collective processes and the nonlinear dynamics of intense charged
particle beam systems. While there has been considerable progress
in three-dimensional numerical and analytical investigations of the
nonlinear Vlasov-Maxwell equations describing intense beam propagation,
there is also considerable interest in the development and application
of simplified one-dimensional kinetic models to describe the longitudinal
nonlinear dynamics of long coasting beams \cite{7hofmann1982,8hofmann1985suppression,9fedele1993thermal,10spentzouris1996direct,11boine1999simulation,12schamel1997theory,13schamel2000kinetic,14davidson2004self,15davidson2004korteweg}
in linear (linac) or large-major-radius ring geometries. The present
paper employs the one-dimensional kinetic formalism recently developed
by Davidson and Startsev \cite{14davidson2004self} for a long coasting
beam propagating through a perfecting conducting circular pipe with
radius $r_{w}$. In Ref.\,\cite{14davidson2004self} the average
longitudinal electric field is expressed as $\left\langle E_{z}\right\rangle \left(z,t\right)=-\left(\partial/\partial z\right)\left\langle \phi\right\rangle \left(z,t\right)=-e_{b}g_{0}\partial\lambda_{b}/\partial z-e_{b}g_{2}r_{w}^{2}\partial^{3}/\partial z^{3}$,
where $e_{b}$ is the particle charge, $g_{0}$ and $g_{z}$ are constant
geometric factors that depend on the location of the conducting wall
and the shape of the transverse density profile, and $\lambda_{b}\left(z,t\right)=\int dp_{z}\,F_{b}\left(z,p_{z},t\right)$
is the line density. In a previous application of the 1D kinetic formalism
developed in Ref.\,\cite{14davidson2004self}, the analyses in Ref.\,\cite{15davidson2004korteweg}
assumed that the longitudinal distribution $F_{b}\left(z,p_{z},t\right)$
corresponded to a so-called waterbag distribution \cite{16davidson2002kinetic,17davidson1972methods,18roberts1967nonlinear,19hohl1967numerical},
where $F_{b}=const.$ within moving boundaries in the phase space
$\left(z,p_{z}\right)$. The weakly nonlinear analysis in Ref.\,\cite{15davidson2004korteweg}
showed that disturbances moving near the sound speed evolve according
to the Korteweg-deVries (KdV) equation \cite{20kordeweg1895change,21washimi1966propagation,22gardner1967method,23ott1969nonlinear,24SeeMethod1}.
The classical KdV equation, which arises in several areas of nonlinear
physics in which there are cubic dispersive corrections to sound-wave-like
signal propagation, also has the appealing feature that it's exactly
solvable using inverse scattering techniques.

While the analysis in Ref.\,\cite{15davidson2004korteweg} reveals
many interesting properties of the nonlinear evolution of longitudinal
disturbances in intense charged particle beams, it is restricted to
the weakly nonlinear regime. In the present analysis, we remove the
restriction to the weakly nonlinear regime, and make use of the 1D
kinetic model developed in Ref.\,\cite{14davidson2004self}, allowing
for moderate to large-amplitude modulation in the charge density of
the beam particles. The organization of this paper is the following.
In Sec.\,\ref{sec: Theory}, the 1D kinetic model \cite{14davidson2004self}
is briefly reviewed (Sec.\,\ref{sub:Theoretical-Model-and}), and
exact (local and nonlocal) nonlinear conservation constraints are
derived (Sec.\,\ref{sub:Conservation-Relations}) for the conservation
of particle number, momentum, and energy per unit length of the beam,
making use of the nonlinear Vlasov equation for $F_{b}\left(z,p_{z},t\right)$
in Eq.\,(\ref{eq:1}), and the expression for $\left\langle E\right\rangle \left(z,t\right)$
in Eq.\,(\ref{eq:2}). Removing the assumption of weak nonlinearity
made in Ref.\,\cite{15davidson2004korteweg}, Sec.\,\ref{sec:Waterbag}
focuses on use of the fully nonlinear kinetic waterbag model (Sec.\,\ref{sub:Kinetic-Waterbag-Model})
to investigate detailed properties of nonlinear pulse-like (soliton)
or periodic traveling-wave disturbance propagating with constant normalized
velocity $M=const.$ relative to the beam frame (Sec.\,\ref{sub:Coherent-Nonlinear-Traveling}).
In normalized variables, $Z'=Z-MT$ and $T'=T$, the waveform of the
disturbance is assumed to be time-stationary ($\partial/\partial T'=0$)
in the frame moving with velocity $M=const.$ relative to the beam
frame. Nonlinear solutions are examined over a wide range of system
parameters, including regimes where the modulation in beam line density
$\lambda_{b}$ exceeds 50\%, corresponding to a strongly bunched beam.
Finally, in Sec.\,\ref{sec:Kinetic} we examine the kinetic model
based on Eqs.\,(\ref{eq:9}) and (\ref{eq:10}) {[}equivalent to
Eqs.\,(\ref{eq:1}) and (\ref{eq:2}){]} for an even broader class
of distribution functions $F_{b}\left(z,p_{z},t\right)$, recognizing
that Eqs.\,(\ref{eq:9}) and (\ref{eq:10}) are Galilean invariant.
{[}Keep in mind that the variables ($z,p_{z},t$) are in the beam
frame, where the particle motion is assumed to be nonrelativistic.{]}
Introducing the appropriately scaled variables (see Sec.\,\ref{sec:Kinetic})
$Z'=Z-MT$, $V_{z}'=V_{z}-M$, $T'=T$, where $M=const.$, we transform
Eqs.\,(\ref{eq:9}) and (\ref{eq:10}) to primed variables, and look
for solutions that are time stationary ($\partial/\partial T'=0$)
in the frame moving with velocity $M=const.$ relative to the beam
frame. The analysis in Sec.\,\ref{sec:Kinetic} parallels the original
Bernstein-Greene-Kruskal (BGK) formulation of BGK solutions to the
1D Vlasov-Poisson equations \cite{25bernstein1957exact,26SeeMethod2},
except for the fact that Eq.\,\eqref{eq:10}, which connects the
effective potential $\left\langle \phi\right\rangle \left(z,t\right)$
to the line density $\lambda_{b}\left(z,t\right)$, has a very different
structure than the 1D Poisson's equation used in the original BGK
analysis. Depending on the choices of trapped-particle and untrapped-particle
distribution functions, the kinetic model described in Sec.\,\ref{sec:Kinetic}
supports a broad range of nonlinear pulse-like (soliton) solutions
and periodic tranveling-wave solution that have stationary waveform
in the frame moving with velocity $M=const.$ relative to the beam
frame. Similar to Sec.\,\ref{sub:Coherent-Nonlinear-Traveling},
the modulation on beam line density can have large amplitude, corresponding
to a strong bunching of the beam particles. Specific examples are
presented in Sec.\,\ref{sec:Kinetic} corresponding to nonlinear
periodic traveling wave solutions.

\section{Theoretical model and assumptions\label{sec: Theory}}

This section provides a brief summary of the one-dimensional kinetic
g-factor model (Sec.\,\ref{sub:Theoretical-Model-and}) developed
by Davidson and Startsev \cite{14davidson2004self} to describe the
nonlinear longitudinal dynamics of a long coasting beam propagating
in the z-direction through a circular, perfectly conducting pipe with
radius $r_{w}$. The 1D kinetic Vlasov equation for the distribution
function $F_{b}\left(z,p_{z},t\right)$ is used (Sec.\,\ref{sub:Conservation-Relations})
to derive several important conservation laws (both local and global)
corresponding to conservation of particle number, momentum, and energy
per unit length of the charge bunch. The results in Secs.\,\ref{sub:Theoretical-Model-and}
and \ref{sub:Conservation-Relations} form the basis for the nonlinear
traveling-wave and traveling-pulse solutions studied in Secs. \ref{sec:Waterbag}
and \ref{sec:Kinetic}.

\subsection{Theoretical Model and Assumptions\label{sub:Theoretical-Model-and}}

This paper makes use of a one-dimensional kinetic model \cite{14davidson2004self}
that describes the nonlinear dynamics of the longitudinal distribution
function $F_{b}\left(z,p_{z},t\right)$, the average self-generated
axial electric field $\langle E_{z}\rangle\left(z,t\right)$, and
the line density $\lambda_{b}\left(z,t\right)=\int dp_{z}\,F_{b}\left(z,p_{z},t\right)$
, for an intense charged particle beam propagating in the z-direction
through a circular, perfectly conducting pipe with radius $r_{w}$.
For simplicity, the analysis is carried out in the beam frame, where
the longitudinal particle motion in $\left(z,p_{z}\right)$ phase
space is assumed to be nonrelativistic, and the beam intensity is
assumed to be sufficiently low that the beam edge radius $r_{b}$
and rms radius $R_{b}=\left\langle r^{2}\right\rangle ^{1/2}=\left\langle x^{2}+y^{2}\right\rangle ^{1/2}$
have a negligibly small dependence on line density $\lambda_{b}$.
Furthermore, properties such as the number density $n_{b}\left(r,z,t\right)$
of beam particles are assumed to be azimuthally symmetric about the
beam axis ($\partial/\partial\theta=0$), where $x=r\cos\theta$ and
$y=r\sin\theta$ are cylindrical polar coordinates. Finally, the axial
spatial variation in the line density $\lambda_{b}\left(z,t\right)=2\pi\int_{0}^{r_{w}}dr\,rn_{b}\left(r,z,t\right)$
is assumed to be sufficiently slow that $k_{z}^{2}r_{w}^{2}\ll1$,
where $\partial/\partial z\backsim k_{z}\backsim L_{z}^{-1}$ is the
inverse length scale of the z-variation.

Making use of these assumptions, it can be shown that the one-dimensional
kinetic equation describing the nonlinear evolution of the longitudinal
distribution function $F_{b}\left(z,p_{z},t\right)$ and average longitudinal
electric field $\langle E_{z}\rangle\left(z,t\right)$ can be expressed
in the beam frame correct to order $k_{z}^{2}r_{w}^{2}$ as \cite{14davidson2004self}

\begin{equation}
\frac{\partial}{\partial t}F_{b}+v_{z}\frac{\partial}{\partial z}F_{b}+e_{b}\left\langle E_{z}\right\rangle \frac{\partial}{\partial p_{z}}F_{b}=0\,,\label{eq:1}
\end{equation}
and

\begin{equation}
\frac{e_{b}}{m_{b}}\left\langle E_{z}\right\rangle =-\frac{U_{b0}^{2}}{\lambda_{b0}}\frac{\partial}{\partial z}\lambda_{b}-\frac{r_{w}^{2}U_{b2}^{2}}{\lambda_{b0}}\frac{\partial^{3}\lambda_{b}}{\partial z^{3}}\,,\label{eq:2}
\end{equation}
Here, $e_{b}$ and $m_{b}$ are the charge and rest mass of a beam
particle, and $\lambda_{b0}=const.$ is a measure of the characteristic
line density of beam particles, e.g.\,, its average value. Moreover,
the constants $U_{b0}^{2}$ and $U_{b2}^{2}$ have dimensions of speed-square,
and are defined by

\begin{equation}
U_{b0}^{2}=\frac{\lambda_{b0}g_{0}e_{b}^{2}}{m_{b}},\thinspace\thinspace\thinspace U_{b2}^{2}=\frac{\lambda_{b0}g_{2}e_{b}^{2}}{m_{b}}\,,\label{eq:3}
\end{equation}
where $g_{0}$ and $g_{2}$ are the geometric factors defined by \cite{14davidson2004self}

\begin{equation}
g_{0}=2\int_{0}^{r_{w}}\frac{dr}{r}\left(2\pi\int_{0}^{r}dr\,r\frac{n_{b}}{\lambda_{b}}\right)^{2}\,,\label{eq:4}
\end{equation}

\begin{equation}
g_{2}=\frac{2}{r_{w}^{2}}\int_{0}^{r_{w}}\frac{dr}{r}\,2\pi\left(\int_{0}^{r}dr\,r\frac{n_{b}}{\lambda_{b}}\right)\int_{0}^{r}dr\,r\int_{r}^{r_{w}}\frac{dr}{r}\left(2\pi\int_{0}^{r}dr\,r\frac{n_{b}}{\lambda_{b}}\right)\,.\label{eq:5}
\end{equation}
In obtaining Eqs.\,(\ref{eq:1})-(\ref{eq:5}), a perfectly conducting
cylindrical wall with $E_{z}\left(r=r_{w},z,t\right)=0$ has been
assumed.

For purposes of illustration, we consider the class of axisymmetric
density profiles $n_{b}\left(r,z,t\right)$ of the form 

\begin{equation}
n_{b}=\begin{cases}
\frac{\lambda_{b}}{\pi r_{b}^{2}}f\left(\frac{r}{r_{b}}\right), & 0\leqslant r<r_{b}\,,\\
0, & r_{b}<r\leqslant r_{w}\,.
\end{cases}\label{eq:6}
\end{equation}
Here, $\lambda_{b}=\int dp_{z}\,F_{b}\left(z,p_{z},t\right)=2\pi\int_{0}^{r_{w}}dr\,rn_{b}\left(r,z,t\right)$
is the line density, $r_{b}$ is the beam edge radius, assumed independent
of $\lambda_{b}$, and $f\left(r/r_{b}\right)$ is the profile shape
function with normalization $\int_{0}^{1}dx\,xf\left(x\right)=1/2$.
As an example, for $f\left(r/r_{b}\right)=\left(n+1\right)\left(1-r^{2}/r_{b}^{2}\right)^{n},\ n=0,1,2,\cdots$,
over the interval $0\leqslant r<r_{b}$, it can be shown that \cite{14davidson2004self}

\begin{equation}
g_{0}=\ln\left(\frac{r_{w}^{2}}{r_{b}^{2}}\right)+\sum_{m=1}^{n+1}\frac{n+1}{m\left(m+n+1\right)}\,,\label{eq:7}
\end{equation}

\begin{equation}
g_{2}=\frac{1}{2}\left[1-\frac{1}{(n+2)}\frac{r_{b}^{2}}{r{}_{w}^{2}}\left(1+\ln\frac{r_{w}^{2}}{r_{b^{2}}}\right)-\sum_{m=1}^{n+1}\frac{1}{m\left(m+n+2\right)}\frac{r_{b}^{2}}{r_{w}^{2}}\right]\,.\label{eq:8}
\end{equation}
From Eqs.\,(\ref{eq:6})-(\ref{eq:8}), we note that $n=0$ corresponds
to a step-function density profile; $n=1$ corresponds to a parabolic
density profile; $n\geqslant2$ corresponds to an even more sharply
peaked density profile; and that the precise values of $g_{0}$ and
$g_{2}$ exhibit a sensitive dependence on profile shape \cite{14davidson2004self}.
Finally, for the choice of shape function $f\left(r/r_{b}\right)=\left(n+1\right)\left(1-r^{2}/r_{b}^{2}\right)^{n},\ n=0,1,2,\cdots$,
it is readily shown that the mean-square beam radius is $R_{b}^{2}=\lambda_{b}^{-1}2\pi\int_{0}^{r_{w}}dr\,r\,r^{2}n_{b}=\left(n+2\right)^{-1}r_{b}^{2}$.

\subsection{Conservation Relations\label{sub:Conservation-Relations}}

Equations (\ref{eq:1}) and (\ref{eq:2}) possess several important
conservation laws, both local and global, corresponding to conservation
of particle number, momentum, and energy per unit length. For present
purposes we express $\left\langle E_{z}\right\rangle \left(z,t\right)=-\left(\partial/\partial z\right)\left\langle \phi\right\rangle \left(z,t\right)$.
Equations (\ref{eq:1}) and (\ref{eq:2}) then describe the evolution
of $F_{b}\left(z,p_{z},t\right)$ and $\langle\phi\rangle\left(z,t\right)$
according to 

\begin{equation}
\frac{\partial}{\partial t}F_{b}+v_{z}\frac{\partial}{\partial z}F_{b}-e_{b}\frac{\partial\left\langle \phi\right\rangle }{\partial z}\frac{\partial F_{b}}{\partial p_{z}}=0\,,\label{eq:9}
\end{equation}
where $v_{z}=p_{z}/m_{b}$ and

\begin{equation}
e_{b}\frac{\partial}{\partial z}\left\langle \phi\right\rangle =m_{b}U_{b0}^{2}\frac{\partial}{\partial z}N_{b}+m_{b}U_{b2}^{2}r_{w}^{2}\frac{\partial^{3}}{\partial z^{3}}N_{b}\,.\label{eq:10}
\end{equation}
Here,

\begin{equation}
N_{b}\left(z,t\right)=\frac{\lambda_{b}\left(z,t\right)}{\lambda_{b0}}=\lambda_{b0}^{-1}\int dp_{z}\,F_{b}\left(z,p_{z},t\right)\label{eq:11}
\end{equation}
is a dimensionles measure of the line density $\lambda_{b}\left(z,t\right)$,
and $\lambda_{b0}=const.$ is the characteristic (e.g. average) value
of line density.

It is convenient to introduce the macroscopic moments

\begin{equation}
N_{b}V_{b}=N_{b}\left\langle v_{z}\right\rangle =\lambda_{b0}^{-1}\int dp_{z}\,v_{z}F_{b}\,,\label{eq:12}
\end{equation}

\begin{equation}
N_{b}\left\langle v_{z}^{n}\right\rangle =\lambda_{b0}^{-1}\int dp_{z}\,v_{z}^{n}F_{b}\,,\label{eq:13}
\end{equation}
where $N_{b}=\lambda_{b}/\lambda_{b0}$ is defined in Eq.\,(\ref{eq:11}),
and $V_{b}\left(z,t\right)=\left(\int dp_{z}\,v_{z}F_{b}\right)\left(\int dp_{z}\,F_{b}\right)$
is the average axial flow velocity in the beam frame. Note that the
effective 'pressure' $P_{b}\left(z,t\right)$ and 'heat flow' $Q_{b}\left(z,t\right)$
are defined (relative to the average flow velocity $V_{b}$) by

\begin{equation}
P_{b}\left(z,t\right)=\lambda_{b0}N_{b}m_{b}\left\langle \left(v_{z}-V_{b}\right)^{2}\right\rangle =m_{b}\int dp_{z}\,\left(v_{z}-V_{b}\right)^{2}F_{b}\,,\label{eq:14}
\end{equation}
and

\begin{equation}
Q_{b}\left(z,t\right)=\lambda_{b0}N_{b}m_{b}\left\langle \left(v_{z}-V_{b}\right)^{3}\right\rangle =m_{b}\int dp_{z}\,\left(v_{z}-V_{b}\right)^{3}F_{b}\,,\label{eq:15}
\end{equation}
where $V_{b}\left(z,t\right)$ is the average flow velocity defined
in Eq.\,(\ref{eq:12}).

We now make use of Eqs.\,(\ref{eq:9}) and (\ref{eq:10}) to derive
the local and global conservation laws corresponding to the conservation
of particle number, momentum, and energy per unit length of the beam.
The subsquent analysis applies to the two classes of beam systems:
(a) a very long, finite-length charge bunch ($L_{b}\gg r_{w}$) with
$N_{b}\left(z\to\pm\infty,t\right)=0$; and (b) a circulating beam
in a large-aspect-ratio ($R_{0}\gg r_{w}$) ring with periodic boundary
condition $N_{b}\left(z+2\pi R_{0},t\right)=N_{b}\left(z,t\right)$
as the beam circulates around the ring with major raduis $R_{0}$.
(Here, $z$ can be viewed as the arc length around the perimiter of
the ring with large radius $R_{0}$.)

\paragraph{Number Conservation:}

From Eqs.\,(\ref{eq:9}), (\ref{eq:11}) and (\ref{eq:12}), operating
on Eq.\,(\ref{eq:9}) with $\lambda_{b0}^{-1}\int dp_{z}\cdots$,
and integrating by parts with respect to $p_{z}$, we obtain

\begin{equation}
\frac{\partial}{\partial t}N_{b}+\frac{\partial}{\partial z}\left(N_{b}V_{b}\right)=0\,,\label{eq:16}
\end{equation}
where $N_{b}\left(z,t\right)=\lambda_{b}\left(z,t\right)/\lambda_{b0}$
is the normalized line density, and $V_{b}\left(z,t\right)$ is the
axial flow velocity {[}Eqs.\,(\ref{eq:11}) and (\ref{eq:12}){]}.
Equation (\ref{eq:16}) is a statement of local number conservation,
i.e.\,, the time rate of change of the local density, $\partial N_{b}/\partial t$,
is equal to minus the derivative of the local flux of particles, $-\left(\partial/\partial z\right)\left(N_{b}V_{b}\right)$.
If we integrate Eq.\,(\ref{eq:16}) over $z$, applying the boundary
conditions described earlier in this section, we obtain

\begin{equation}
\frac{\partial}{\partial t}\int dzN_{b}=0\,,\label{eq:17}
\end{equation}
which corresponds to the global conservation of the number of beam
particles.

\paragraph{Momentum Conservation:}

We now operate on Eq.\,(\ref{eq:9}) with $\lambda_{b0}^{-1}\int dp_{z}\,p_{z}\cdots$,
where $p_{z}=m_{b}v_{z}$, and make use of Eqs.\,(\ref{eq:12}) and
(\ref{eq:13}). This gives

\begin{equation}
\frac{\partial}{\partial t}N_{b}m_{b}V_{b}+\frac{\partial}{\partial z}N_{b}m_{b}\left\langle v_{z}v_{z}\right\rangle +e_{b}N_{b}\frac{\partial}{\partial z}\left\langle \phi\right\rangle =0\,,\label{eq:18}
\end{equation}
where $-\left(\partial/\partial z\right)\left(\left\langle \phi\right\rangle \right)$
is defined in Eq.\,(\ref{eq:10}), and we have integrated by parts
with respect to $p_{z}$ to obtain Eq.\,(\ref{eq:18}) from Eq.\,(\ref{eq:9}).
Equation (\ref{eq:18}) can be expressed in an alternate form by making
use of Eq.\,(\ref{eq:10}) to eliminate $e_{b}\left(\partial/\partial z\right)\left\langle \phi\right\rangle $
and combine Eqs.\,(\ref{eq:12})-(\ref{eq:14}) to express 

\begin{equation}
N_{b}m_{b}\left\langle v_{z}v_{z}\right\rangle =N_{b}m_{b}V_{b}V_{b}+\lambda_{b0}^{-1}P_{b}\,,\label{eq:19}
\end{equation}
where $V_{b}\left(z,t\right)$ is the average flow velocity, and $P_{b}\left(z,t\right)$
is the effective pressure of the beam particles. Substituting Eqs.\,(\ref{eq:10})
and (\ref{eq:19}) into Eq.\,(\ref{eq:18}), we obtain

\begin{align}
 & \frac{\partial}{\partial t}N_{b}m_{b}V_{b}+\frac{\partial}{\partial z}\left\{ N_{b}m_{b}V_{b}V_{b}+\lambda_{b0}^{-1}P_{b}\right\} +N_{b}m_{b}\left\{ U_{b0}^{2}\frac{\partial N_{b}}{\partial z}+U_{b2}^{2}r_{w}^{2}\frac{\partial^{2}N_{b}}{\partial z^{2}}\right\} \nonumber \\
 & =\frac{\partial}{\partial t}N_{b}m_{b}V_{b}+\frac{\partial}{\partial z}\left\{ N_{b}m_{b}V_{b}V_{b}+\lambda_{b0}^{-1}P_{b}+\frac{1}{2}m_{b}U_{b0}^{2}N_{b}^{2}\right.\nonumber \\
 & \left.+\frac{\partial}{\partial z}m_{b}U_{b2}^{2}r_{w}^{2}\left[N_{b}\frac{\partial^{2}N_{b}}{\partial z^{2}}-\frac{1}{2}\left(\frac{\partial N_{b}}{\partial z}\right)^{2}\right]\right\} =0\,.\label{eq:20}
\end{align}
Note that Eq.\,(\ref{eq:20}) expresses the local force balance equation
in the form of a local conservation relation for the momentum density
of a beam fluid element. Moreover, integrating Eq.\,(\ref{eq:20})
over $z$ and applying the boundary conditions described earlier in
Sec.\,\ref{sec: Theory} gives

\begin{equation}
\frac{\partial}{\partial t}\int dz\,N_{b}m_{b}V_{b}=0\,,\label{eq:21}
\end{equation}
which corresponds to global momentum conservation.

\paragraph{Energy Conservation:}

We now operate on Eq.\,(\ref{eq:9}) with $\lambda_{b0}^{-1}\int dp_{z}\,\frac{1}{2}m_{b}v_{z}^{2}\cdots$
and make use of Eqs.\,(\ref{eq:11})-(\ref{eq:13}) and $p_{z}=m_{b}v_{z}$.
Integrating by parts with respect to $p_{z}$, we readily obtain

\begin{equation}
\frac{\partial}{\partial z}\frac{1}{2}N_{b}m_{b}\left\langle v_{z}^{2}\right\rangle +\frac{\partial}{\partial z}\frac{1}{2}N_{b}m_{b}\left\langle v_{z}^{3}\right\rangle +e_{b}\frac{\partial\left\langle \phi\right\rangle }{\partial z}N_{b}V_{b}=0\,.\label{eq:22}
\end{equation}
From Eqs.\,(\ref{eq:10}) and (\ref{eq:16}), some straightforward
algebraic manipulation gives

\begin{align}
e_{b}\frac{\partial\left\langle \phi\right\rangle }{\partial z}N_{b}V_{b} & =\frac{\partial}{\partial z}\left[\frac{1}{2}m_{b}U_{b0}^{2}N_{b}^{2}-\frac{1}{2}m_{b}U_{b2}^{2}r_{w}^{2}\left(\frac{\partial N_{b}}{\partial z}\right)^{2}\right]\nonumber \\
 & +\frac{\partial}{\partial z}\left\{ m_{b}U_{b0}^{2}\left(N_{b}N_{b}V_{b}\right)+m_{b}U_{b2}^{2}r_{w}^{2}\left[N_{b}V_{b}\frac{\partial^{2}N_{b}}{\partial z^{2}}+\frac{\partial N_{b}}{\partial z}\frac{\partial N_{b}}{\partial t}\right]\right\} \,.\label{eq:23}
\end{align}
Substituting Eq.\,(\ref{eq:23}) into Eq.\,(\ref{eq:22}) and rearranging
terms, we obtain

\begin{align}
\frac{\partial}{\partial t}\left\{ \frac{1}{2}N_{b}m_{b}\left\langle v_{z}^{2}\right\rangle +\frac{1}{2}m_{b}U_{b0}^{2}N_{b}^{2}-\frac{1}{2}m_{b}U_{b2}^{2}r_{w}^{2}\left(\frac{\partial N_{b}}{\partial z}\right)^{2}\right\} +\nonumber \\
\frac{\partial}{\partial z}\left\{ \frac{1}{2}N_{b}m_{b}\left\langle v_{z}^{3}\right\rangle +m_{b}U_{b0}^{2}N_{b}^{2}V_{b}+m_{b}U_{b2}^{2}r_{w}^{2}\left[N_{b}V_{b}\frac{\partial^{2}N_{b}}{\partial z^{2}}+\frac{\partial N_{b}}{\partial z}\frac{\partial N_{b}}{\partial t}\right]\right\}  & =0\,,\label{eq:24}
\end{align}
which corresponds to local conservation of energy. Global energy conservation
follows upon integrating Eq.\,(\ref{eq:24}) over $z$, which gives

\begin{equation}
\frac{\partial}{\partial t}\int dz\left\{ \frac{1}{2}N_{b}m_{b}\left\langle v_{z}^{2}\right\rangle +\frac{1}{2}m_{b}U_{b0}^{2}N_{b}^{2}-\frac{1}{2}m_{b}U_{b2}^{2}r_{w}^{2}\left(\frac{\partial N_{b}}{\partial z}\right)^{2}\right\} =0\,.\label{eq:25}
\end{equation}
Note that Eq.\,(\ref{eq:25}) describes the balance in energy exchange
between particle kinetic energy and electrostatic field energy. Moreover,
the final two terms in Eq.\,(\ref{eq:25}) correspond to electrostatic
field energy, and the term proportional to $U_{b0}^{2}$ is positive,
whereas the term proportional to $U_{b2}^{2}$ is manifestly negative.
Because of the negative sign of the third term in Eq.\,(\ref{eq:25}),
note that any increase in $\left(\partial N_{b}/\partial z\right)^{2}$
averaged over $z$ must compensated by a corresponding increase in
the first two terms in Eq.\,(\ref{eq:25}).

To summarize, the local conservation laws in Eqs.\,(\ref{eq:16}),
(\ref{eq:18}) and (\ref{eq:24}), and the global conservation laws
in Eqs.\,(\ref{eq:17}), (\ref{eq:21}) and (\ref{eq:25}), provide
powerful nonlinear constrains on the evolution of the normalized line
density $N_{b}$, momentum density $N_{b}m_{b}V_{b}$, and kinetic
energy density $N_{b}m_{b}\left\langle v_{z}^{2}\right\rangle /2=N_{b}m_{b}V_{b}^{2}/2+\lambda_{b0}^{-1}P_{b}/2$.
Furthermore, these conservation constraints are exact consequences
of the 1D nonlinear Vlasov equation (\ref{eq:9}) for $F_{b}\left(z,p_{z},t\right)$,
where $e_{b}\left(\partial/\partial z\right)\left\langle \phi\right\rangle \left(z,t\right)$
is defined in Eq.\,(\ref{eq:10}), and $U_{b0}^{2}$ and $U_{b2}^{2}$
are expressed in terms of the geometric factors $g_{0}$ and $g_{2}$
in Eq.\,(\ref{eq:3}).

Finally, the energy balance equation (\ref{eq:22}), the momentum
balance equation (\ref{eq:18}), and the continuity equation (\ref{eq:16})
can be combined to give a dynamical equation for the evolution of
the effective pressure $P_{b}\left(z,t\right)$ of the beam particles.
We make use of $N_{b}m_{b}\left\langle v_{z}^{2}\right\rangle =N_{b}m_{b}V_{b}^{2}+\lambda_{b0}^{-1}P_{b}$,
and express

\begin{align}
N_{b}m_{b}\left\langle v_{z}^{3}\right\rangle  & =N_{b}m_{b}\left\langle \left(v_{z}-V_{b}+V_{b}\right)^{3}\right\rangle \nonumber \\
 & =N_{b}m_{b}V_{b}^{3}+3N_{b}m_{b}V_{b}\left\langle \left(v_{z}-V_{b}\right)^{2}\right\rangle +N_{b}m_{b}\left\langle \left(v_{z}-V_{b}\right)^{3}\right\rangle \nonumber \\
 & =N_{b}m_{b}V_{b}^{3}+3N_{b}V_{b}\lambda_{b0}^{-1}P_{b}+\lambda_{b0}^{-1}Q_{b}\,,\label{eq:26}
\end{align}
where $Q_{b}$ is the effective heat flow defined in Eq.\,(\ref{eq:15}).
Without presenting algebraic details, some straightforward manipulation
of Eq.\,(\ref{eq:22}) that make use of Eqs.\,(\ref{eq:16}) and
(\ref{eq:18}) then gives

\begin{equation}
\left(\frac{\partial}{\partial t}+V_{b}\frac{\partial}{\partial z}\right)P_{b}+3P_{b}\frac{\partial V_{b}}{\partial z}+\frac{\partial}{\partial z}Q_{b}=0\,.\label{eq:27}
\end{equation}

To summarize, Eqs.\,(\ref{eq:16}), (\ref{eq:20}) and (\ref{eq:27})
describe the self-consistent nonlinear of $N_{b}\left(z,t\right)$,
$V_{b}\left(z,t\right)$ and $P_{b}\left(z,t\right)$. In the special
case where the heat flow contribution $\left(\partial/\partial z\right)Q_{b}$
is negligibly small in Eq.\,(\ref{eq:27}), the pressure $P_{b}\left(z,t\right)$
evolves approximately according to 

\begin{equation}
\left(\frac{\partial}{\partial t}+V_{b}\frac{\partial}{\partial z}\right)P_{b}+3P_{b}\frac{\partial V_{b}}{\partial z}=0\,.\label{eq:28}
\end{equation}
The continuity equation (\ref{eq:16}) can be expressed as

\begin{equation}
\left(\frac{\partial}{\partial t}+V_{b}\frac{\partial}{\partial z}\right)N_{b}+N_{b}\frac{\partial V_{b}}{\partial z}=0\,.\label{eq:29}
\end{equation}
Combining Eqs.\,(\ref{eq:27}) and (\ref{eq:28}) , we obtain 

\begin{equation}
\left(\frac{\partial}{\partial t}+V_{b}\frac{\partial}{\partial z}\right)\left(\frac{P_{b}}{N_{b}^{3}}\right)=0\,,\label{eq:30}
\end{equation}
which can be integrated to give the triple-adiabatic pressure relation
$\left(P_{b}/N_{b}^{3}\right)=const.$ Therefore, for negligibly small
heat flow in Eq.\,(\ref{eq:27}), the macroscopic fluid model obtained
by taking moments of the 1D Vlasov equation (\ref{eq:9}) closes,
and the nonlinear evolution of $N_{b}$, $V_{b}$ and $P_{b}$ is
described by Eqs.\,(\ref{eq:16}), (\ref{eq:20}) and (\ref{eq:30}).

In Sec.\,\ref{sec:Waterbag}, we discuss a particular choice of distribution
function $F_{b}\left(z,p_{z},t\right)$, corresponding to the so-called
waterbag distribution, for which the heat flow $Q_{b}\left(z,t\right)$
is exactly zero during the nonlinear evolution of the system. In this
case, the closure is exact, and the nonlinear evolution of the system
is fully described by Eqs.\,(\ref{eq:16}), (\ref{eq:20}) and (\ref{eq:30}).

\section{Coherent sonlinear structures obtained from kinetic waterbag model
\label{sec:Waterbag}}

The 1D kinetic g-factor model based on Eqs.\,(\ref{eq:1}) and (\ref{eq:2})
can be used to determine the nonlinear evolution of the beam distribution
function $F_{b}\left(z,p_{z},t\right)$ for a broad range of system
parameters and initial distribution functions. In this section, we
examine Eqs.\,(\ref{eq:1}) and (\ref{eq:2}) for the class of exact
solutions for $F_{b}\left(z,p_{z},t\right)$ corresponding to the
so-called waterbag distribution in which $F_{b}\left(z,p_{z},t\right)$
has uniform density in phase space (Sec.\,\ref{sub:Kinetic-Waterbag-Model}).
The subclass of coherent nonlinear traveling-wave and traveling-pulse
solutions with undistorted waveform are then examined (Sec.\,\ref{sub:Coherent-Nonlinear-Traveling})
for disturbances traveling in the longitudinal direction with constant
normalized velocity $M=const.$

\subsection{Kinetic Waterbag Model\label{sub:Kinetic-Waterbag-Model}}

Equations (\ref{eq:1}) and (\ref{eq:2}), or equivalently, Eqs.\,(\ref{eq:9})
and (\ref{eq:10}) constitute the starting point in the present 1D
kinetic description of the longitudinal nonlinear dynamics of a long
coasting beam. The detailed wave excitations associated with Eqs.\,(\ref{eq:9})
and (\ref{eq:10}) of course depend on the form of the distribution
function $F_{b}\left(z,p_{z},t\right)$. For small-amplitude perturbations,
Eqs.\,(\ref{eq:1}) and (\ref{eq:2}) support solutions corresponding
to sound-wave-like disturbances with signal speed depending on $U_{b0}$
and the momentum spread of $F_{b}$, and cubic dispersive modifications
depending on $U_{b2}$ \cite{14davidson2004self}.

In this section, we specialize to the class of exact nonlinear solutions
for $F_{b}\left(z,p_{z},t\right)$ to Eq.\,(\ref{eq:1}) corresponding
to the waterbag distribution \cite{15davidson2004korteweg,16davidson2002kinetic,17davidson1972methods,18roberts1967nonlinear,19hohl1967numerical}

\begin{equation}
F_{b}\left(z,p_{z},t\right)=\begin{cases}
A=const. & -m_{b}V_{b}^{-}\left(z,t\right)<p_{z}<m_{b}V_{b}^{+}\left(z,t\right)\,,\\
0\,, & otherwise\,,
\end{cases}\label{eq:31}
\end{equation}
for $-\infty<z<\infty$ (long coasting beam in linear geometry) or
$0<z<2\pi R_{0}$ (large-aspect-ratio ring with major radius $R_{0}$).
In Eq.\,\eqref{eq:31}, the distribution function $F_{b}=A$ remains
constant within the boundary curves $-m_{b}V_{b}^{-}$ and $+m_{b}V_{b}^{+}$,
and zero outside. The boundary curves $-m_{b}V_{b}^{-}\left(z,t\right)$
and $+m_{b}V_{b}^{+}\left(z,t\right)$, assumed single-valued, of
course distort nonlinearly as the system evolves according to Eqs.\,(\ref{eq:1})
and (\ref{eq:2}) {[}or equivalently, Eqs.\,(\ref{eq:9}) and (\ref{eq:10}){]}.
We integrate across the two boundary curves in Eq.\,\ref{eq:31}
by operating on Eq.\,(\ref{eq:1}) with

\begin{equation}
\lim_{\epsilon\to0^{+}}\int_{m_{b}V_{b}^{-}\left(1-\epsilon\right)}^{m_{b}V_{b}^{-}\left(1+\epsilon\right)}dp_{z}\,p_{z}\cdots\,,\ and\ \lim_{\epsilon\to0^{+}}\int_{m_{b}V_{b}^{+}\left(1-\epsilon\right)}^{m_{b}V_{b}^{+}\left(1+\epsilon\right)}dp_{z}\,p_{z}\cdots\,,\label{eq:32}
\end{equation}
where $p_{z}=m_{b}v_{z}$. Integrating by parts with respect to $p_{z}$,
and taking the limit $\epsilon\to0^{+}$, we obtain for the nonlinear
evolution of the boundary curves $V_{b}^{-}\left(z,t\right)$ and
$V_{b}^{+}\left(z,t\right)$ 

\begin{equation}
\frac{\partial}{\partial t}V_{b}^{-}+V_{b}^{-}\frac{\partial}{\partial z}V_{b}^{-}=\frac{e_{b}}{m_{b}}\left\langle E_{z}\right\rangle \,,\label{eq:33}
\end{equation}

\begin{equation}
\frac{\partial}{\partial t}V_{b}^{+}+V_{b}^{+}\frac{\partial}{\partial z}V_{b}^{+}=\frac{e_{b}}{m_{b}}\left\langle E_{z}\right\rangle \,,\label{eq:34}
\end{equation}
where $\left\langle E_{z}\right\rangle $ is defined in Eq.\,(\ref{eq:2}).

For the choice of waterbag distribution in Eq.\,(\ref{eq:31}), we
calculate several macroscopic fluid quantities {[}see also Eqs.\,(\ref{eq:11})-(\ref{eq:15}){]}
corresponding to line density

\begin{equation}
\lambda_{b}=\int dp_{z}\,F_{b}=Am_{b}\left(V_{b}^{+}-V_{b}^{-}\right)\,,\label{eq:35}
\end{equation}
axial flow velocity

\begin{equation}
V_{b}=\lambda_{b}^{-1}\int dp_{z}\,v_{z}F_{b}=\frac{1}{2}\left(V_{b}^{+}+V_{b}^{-}\right)\,,\label{eq:36}
\end{equation}
beam particle pressure

\begin{equation}
P_{b}=m_{b}\int dp_{z}\,\left(v_{z}-V_{b}\right)^{2}F_{b}=\frac{1}{12}m_{b}A\left(V_{b}^{+}-V_{b}^{-}\right)^{3}=\frac{1}{12\left(m_{b}A\right)^{2}}\lambda_{b}^{3}\,,\label{eq:37}
\end{equation}
and beam particle heat flow

\begin{equation}
Q_{b}=m_{b}\int dp_{z}\left(v_{z}-V_{b}\right)^{3}F_{b}=0\,.\label{eq:38}
\end{equation}
Note that the heat flow is exactly $Q_{b}=0$ for the choice of waterbag
distribution in Eq.\,(\ref{eq:31}).

Making use of the dynamical equations for $V_{b}^{-}\left(z,t\right)$
and $V_{b}^{+}\left(z,t\right)$ in Eqs.\,(\ref{eq:33}) and (\ref{eq:34}),
where $\left\langle E_{z}\right\rangle $ is defined in Eq.\,(\ref{eq:2}),
some straightforward algebra shows that $\lambda_{b}\left(z,t\right)$,
$V_{b}\left(z,t\right)$ and $P_{b}\left(z,t\right)$ evolve according
to 

\begin{equation}
\frac{\partial}{\partial t}\lambda_{b}+\frac{\partial}{\partial z}\left(\lambda_{b}V_{b}\right)=0\,,\label{eq:39}
\end{equation}

\begin{equation}
\lambda_{b}\left(\frac{\partial}{\partial t}V_{b}+V_{b}\frac{\partial}{\partial z}V_{b}\right)+\frac{1}{m}\frac{\partial P_{b}}{\partial z}=-\lambda_{b}\left(\frac{U_{b0}^{2}}{\lambda_{b0}}\frac{\partial}{\partial z}\lambda_{b}+\frac{U_{b2}^{2}r_{w}^{2}}{\lambda_{b0}}\frac{\partial^{3}\lambda_{b}}{\partial z^{3}}\right)\,,\label{eq:40}
\end{equation}

\begin{equation}
\left(\frac{\partial}{\partial t}+V_{b}\frac{\partial}{\partial z}\right)\left(\frac{P_{b}}{\lambda_{b}^{3}}\right)=0\,.\label{eq:41}
\end{equation}
Note from Eqs.\,(\ref{eq:37}) and (\ref{eq:41}) that $P_{b}\left(z,t\right)$
can be expressed as

\begin{equation}
P_{b}\left(z,t\right)=\frac{P_{b0}}{\lambda_{b0}^{3}}\lambda_{b}^{3}\left(z,t\right)\,,\label{eq:42}
\end{equation}
where $P_{b0}=const.$ and $\lambda_{b0}=const.$ represent the characteristic
(e.g.\,, average) value of the pressure and line density of the beam
particles respectively, and $P_{b0}/\lambda_{b0}^{3}=1/12\left(m_{b}A\right)^{2}=const.$\,,
where $A$ is the constant phase-space density in Eq.\,(\ref{eq:31}).
By virtue of the fact that the heat flow $Q_{b}\left(z,t\right)=0$
exactly for the choice of distribution function $F_{b}\left(z,p_{z},t\right)$
in Eq.\,(\ref{eq:31}), it is not surprising that Eqs.\,(\ref{eq:39})-(\ref{eq:42})
are identical to the macroscopic fluid equations (\ref{eq:16}), (\ref{eq:20})
and (\ref{eq:30}), obtained in Sec.\,\ref{sub:Conservation-Relations},
where Eq.\,(\ref{eq:30}) has made the assumption of negligible heat
flow in Eq.\,(\ref{eq:27}). Note here that $N_{b}\left(z,t\right)$
and $\lambda_{b}\left(z,t\right)$ are related by $N_{b}\left(z,t\right)=\lambda_{b}\left(z,t\right)/\lambda_{b0}$.

For present purposes, we introduce the effective thermal speed $U_{bT}$
associated with the waterbag distribution in Eq.\,(\ref{eq:31})
defined by

\begin{equation}
U_{bT}^{2}=\frac{3P_{b0}}{\lambda_{b0}m_{b}}\,,\label{eq:43}
\end{equation}
and the nomalized (dimensionless) fluid quantities $\eta\left(z,t\right)$
and $U\left(z,t\right)$ defined by

\begin{equation}
\eta=N_{b}-1=\frac{\lambda_{b}-\lambda_{b0}}{\lambda_{b0}},\ U=\frac{V_{b}}{\left(U_{b0}^{2}+U_{bT}^{2}\right)^{1/2}}\,.\label{eq:44}
\end{equation}
In Eq.\,(\ref{eq:44}), $\left(U_{b0}^{2}+U_{bT}^{2}\right)^{1/2}$
is the effective sound speed associated with the geometric factor
$g_{0}$ and the thermal speed $U_{bT}$. Furthermore, we introduce
the scaled (dimensionless) time variable $T$ and spatial variable
$Z$ defined by

\begin{equation}
T=\left(\frac{U_{b0}^{2}+U_{bT}^{2}}{U_{b2}^{2}}\right)\frac{U_{b2}t}{r_{w}},\ Z=\left(\frac{U_{b0}^{2}+U_{bT}^{2}}{U_{b2}^{2}}\right)\frac{z}{r_{w}}\,.\label{eq:45}
\end{equation}
Making use of the macroscopic equations (\ref{eq:39}), (\ref{eq:40})
and (\ref{eq:42}), and the definitions in Eqs.\,(\ref{eq:43})-(\ref{eq:45}),
it is straightforward to show that the continuity equation (\ref{eq:39})
and force balance equation (\ref{eq:40}) reduce in dimensionless
variables exactly to 

\begin{equation}
\frac{\partial}{\partial T}\eta+\frac{\partial}{\partial Z}\left(U+\eta U\right)=0,\label{eq:46}
\end{equation}

\begin{equation}
\frac{\partial}{\partial T}U+\frac{\partial}{\partial Z}\left(\frac{1}{2}U^{2}+\eta+\frac{1}{2}\frac{U_{bT}^{2}}{U_{b0}^{2}+U_{bT}^{2}}\eta^{2}+\frac{\partial^{2}}{\partial Z^{2}}\eta\right)=0\,.\label{eq:47}
\end{equation}
The fluid description in scaled variables provided by Eqs.\,(\ref{eq:46})
and (\ref{eq:47}) is exactly equivalent to the kinetic description
provided by Eqs.\,(\ref{eq:1}) and (\ref{eq:2}) for the choice
of waterbag distribution in Eq.\,(\ref{eq:31}).

\subsection{Coherent Nonlinear Traveling Ware and Pulse Solutions\label{sub:Coherent-Nonlinear-Traveling}}

Within the context of the present 1D model, Eqs.\,(\ref{eq:46})
and (\ref{eq:47}) can be used to investigate detailed properties
of collective excitations over a wide range of system parameters.
For example, in the weakly nonlinear regime, for small-amplitude disturbances
moving near the sound speed $\left(U_{b0}^{2}+U_{bT}^{2}\right)^{1/2}$,
Eqs.\,(\ref{eq:46}) and (\ref{eq:47}) can be shown to reduce to
the Korteweg-deVries equation \cite{15davidson2004korteweg}, which
exhibits the generation and interaction of coherent structures (solitons)
for a wide range of initial density pertuibations $\eta\left(Z,T=0\right)\neq0$
\cite{22gardner1967method}. While the analysis in Ref \cite{15davidson2004korteweg}
has several interesting features, the results are limited to the weakly
nonlinear regime where $|\eta|\ll1$ and $|U|\ll1$.

In this paper, we examine Eqs.\,(\ref{eq:46}) and (\ref{eq:47})
in circumstances where there are not \textit{a priori} restrictions
to small amplitude, i.e.\,, $\eta=\left(\lambda_{b}-\lambda_{b0}\right)/\lambda_{b0}$
is allowed to be of order unity, as long as $\lambda_{b}/\lambda_{b0}>0$,
which corresponds to $\eta>-1$. Furthermore, we look for solutions
to Eqs.\,(\ref{eq:46}) and (\ref{eq:47}) that depend on $Z$ and
$T$ exclusively through the variables $Z'=Z-MT$ and $T'=T$, where
$M=const.$ is the normalized pulse speed measured in units of the
sound speed $\left(U_{b0}^{2}+U_{bT}^{2}\right)^{1/2}$. Making use
of $\partial/\partial Z=\partial/\partial Z'$ and $\partial/\partial T=\partial/\partial T'-M\partial/\partial Z'$
and looking for time-stationary solutions ($\partial/\partial T'=0$)
in the frame of reference moving with normalized velocity $M=const.$\,,
Eqs.\,(\ref{eq:46}) and (\ref{eq:47}) for $\eta\left(Z'\right)$
and $U\left(Z'\right)$ become 

\begin{equation}
\frac{\partial}{\partial Z'}\left[\left(-M+U\right)\eta+U\right]=0\,,\label{eq:48}
\end{equation}

\begin{equation}
\frac{\partial}{\partial Z'}\left[\frac{1}{2}U^{2}-MU+\eta+\frac{1}{2}\frac{U_{bT}^{2}}{U_{b0}^{2}+U_{bT}^{2}}\eta^{2}+\frac{\partial^{2}\eta}{\partial Z'^{2}}\right]=0\,.\label{eq:49}
\end{equation}
Integrating with respect to $Z'$, Eqs.\,(\ref{eq:48}) and (\ref{eq:49})
give

\begin{equation}
-M\eta+\left(1+\eta\right)U=const.\,,\label{eq:50}
\end{equation}

\begin{equation}
\frac{1}{2}U^{2}-MU+\eta+\frac{1}{2}\frac{U_{bT}^{2}}{U_{b0}^{2}+U_{bT}^{2}}\eta^{2}+\frac{\partial^{2}\eta}{\partial Z'^{2}}=const.\,,\label{eq:51}
\end{equation}
which relate $\eta\left(Z'\right)$ and $U\left(Z'\right)$, where
$Z'=Z-MT$ . 

The solutions for $\eta\left(Z'\right)$ and $U\left(Z'\right)$ to
Eqs.\,(\ref{eq:50}) and (\ref{eq:51}) depend on the values of the
constants in Eqs.\,(\ref{eq:50}) and (\ref{eq:51}). For present
purposes we consider boundary conditions such that $U=0$ when $\eta=0$,
and $\eta''=0$ when $U=0$ and $\eta=0$. In this case the values
of the constants in Eqs.\,(\ref{eq:50}) and (\ref{eq:51}) are zero,
which gives

\begin{equation}
U=M\frac{\eta}{1+\eta}\,.\label{eq:52}
\end{equation}

\begin{equation}
\frac{\partial^{2}\eta}{\partial Z'^{2}}+\left\{ \frac{1}{2}\left(U-M\right)^{2}-\frac{1}{2}M^{2}+\eta+\frac{1}{2}\frac{U_{bT}^{2}}{U_{b0}^{2}+U_{bT}^{2}}\eta^{2}\right\} =0\,.\label{eq:53}
\end{equation}
Substituting Eq.\,(\ref{eq:52}) into Eq.\,(\ref{eq:53}), we obtain

\begin{equation}
\frac{\partial^{2}\eta}{\partial Z'^{2}}+\left\{ \frac{1}{2}M^{2}\left(\frac{1}{\left(1+\eta\right)^{2}}-1\right)+\eta+\frac{1}{2}\frac{U_{bT}^{2}}{U_{b0}^{2}+U_{bT}^{2}}\eta^{2}\right\} =0\,,\label{eq:54}
\end{equation}
which is a second-order nonlinear differential equation for the perturbation
in line density $\eta\left(Z'\right)=\left[\lambda_{b}\left(Z'\right)-\lambda_{b0}\right]/\lambda_{b0}$,
where $Z'=Z-MT$. Some straightforward algebraic manipulation shows
that Eq.\,(\ref{eq:54}) can be expressed in the equivalent form

\begin{equation}
\frac{\partial^{2}\eta}{\partial Z'^{2}}=-\frac{\partial}{\partial\eta}V\left(\eta\right)\,,\label{eq:55}
\end{equation}
where $V\left(\eta\right)$ is the effective potential defined by

\begin{equation}
V\left(\eta\right)=\frac{1}{2}\frac{\eta^{2}}{1+\eta}\left\{ \epsilon_{T}\eta^{2}+\left(1+\epsilon_{T}\right)\eta+\left(1-M^{2}\right)\right\} \,,\label{eq:56}
\end{equation}
and the dimensionless parameter $\epsilon_{T}$, defined by

\begin{equation}
\epsilon_{T}=\frac{1}{3}\left(\frac{U_{bT}^{2}}{U_{b0}^{2}+U_{bT}^{2}}\right)\,,\label{eq:57}
\end{equation}
is a measure of the longitudinal thermal speed of the beam particle.

Equations (\ref{eq:55}) and (\ref{eq:56}) can be used to determine
the solutions for $\eta\left(Z'\right)$ for a broad range of dimensionless
parameters $\epsilon_{T}$ and $M$. Furthermore, Eqs.\,(\ref{eq:55})
and (\ref{eq:56}) have been obtained from Eqs.\,(\ref{eq:50}) and
(\ref{eq:51}) for the special class of boundary conditions where
$U=0$ and $\eta''=0$ when $\eta=0$ {[}see discussion prior to Eqs.\,(\ref{eq:52})
and (\ref{eq:53}){]}. Indeed, we will show below that Eqs.\,(\ref{eq:55})
and (\ref{eq:56}) support two classses of solutions consistent with
these boundary conditions. These correspond to: (a) localized (pulse-like)
soliton solutions when $M^{2}>1$, satisfying $\eta\left(Z'=\pm\infty\right)=0$,
$U\left(Z'=\pm\infty\right)=0$, and $\left[\partial^{2}\eta/\partial Z'^{2}\right]_{Z'=\pm\infty}=0$;
and (b) nonlinear periodic traveling-wave solutions when $M^{2}<1$,
with $\eta\left(Z'\right)=\eta\left(Z'+L\right)$, and $\eta\left(Z'=0\right)=0$,
$U\left(Z'=0\right)=0$, and $\left[\partial^{2}\eta/\partial Z'^{2}\right]_{Z'=0}=0$.

In general, the effective potential $V\left(\eta\right)$ in Eq.\,(\ref{eq:57})
can be expressed as

\begin{equation}
V\left(\eta\right)=\frac{1}{2}\frac{\eta^{2}}{1+\eta}\epsilon_{T}\left[\left(\eta-\eta^{+}\right)\left(\eta-\eta^{-}\right)\right]\,,\label{eq:58}
\end{equation}
where

\begin{equation}
\eta^{\pm}=\frac{1}{2}\left\{ -\left(1+\frac{1}{\epsilon_{T}}\right)\pm\left[\left(1+\frac{1}{\epsilon_{T}}\right)^{2}+\frac{4}{\epsilon_{T}}\left(M^{2}-1\right)^{1/2}\right]\right\} \,.\label{eq:59}
\end{equation}
In Eqs.\,(\ref{eq:58}) and (\ref{eq:59}), $\epsilon_{T}$ is restricted
to the range $0<\epsilon_{T}<1/3$, and $M^{2}$ can satisfy $M^{2}>1$
or $M^{2}<1$. Examination of Eq.\,(\ref{eq:59}) shows that

\begin{equation}
\eta^{-}<-1,\ \eta^{+}>-1,\label{eq:60}
\end{equation}
for all allowed values of $\epsilon_{T}$ and $M^{2}$. Furthermore,
it's also clear from Eq.\,(\ref{eq:59}) that

\begin{equation}
\begin{cases}
\eta^{+}>0, & for\ M^{2}>1\,,\\
\eta^{+}<0, & for\ M^{2}<1\,.
\end{cases}\label{eq:61}
\end{equation}
Recall that $\eta=\left(\lambda_{b}-\lambda_{b0}\right)/\lambda_{b0}$.
Then $\lambda_{b}/\lambda_{b0}\geqslant0$ implies that $\eta\geqslant-1$
is the region of interest physically for solutions to Eq.\,(\ref{eq:1}).
Note that Eq.\,(\ref{eq:55}) has the form of a dynamical equation
of motion, with $\eta$ playing the role of displacement, $Z'$ playing
the role of time, and $V\left(\eta\right)$ playing the role of effective
potential. Multiplying Eq.\,(\ref{eq:55}) by $\partial\eta/\partial Z'$
and integrating, we obtain

\begin{equation}
\frac{1}{2}\left(\frac{\partial\eta}{\partial Z'}\right)^{2}+V\left(\eta\right)=E=const.\label{eq:62}
\end{equation}
Equation (\ref{eq:62}) plays the role of an energy conservation constraint,
and can be integrated to determine $\eta\left(Z'\right)$ for the
effective potential $V\left(\eta\right)$ defined in Eq.\,(\ref{eq:58}).
We now examine solutions to Eq.\,(\ref{eq:62}) for the two cases
indentified earlier: $M^{2}<1$ and $-1<\eta^{+}<0$; and $M^{2}>1$
and $\eta^{+}>0$.

\paragraph{Nonlinear Traveling-wave Solutions\,$\left(M^{2}<1\ and\ -1<\eta^{+}<0\right)$:}

\begin{figure}[ptb]
\begin{centering}
\includegraphics[width=3.5in]{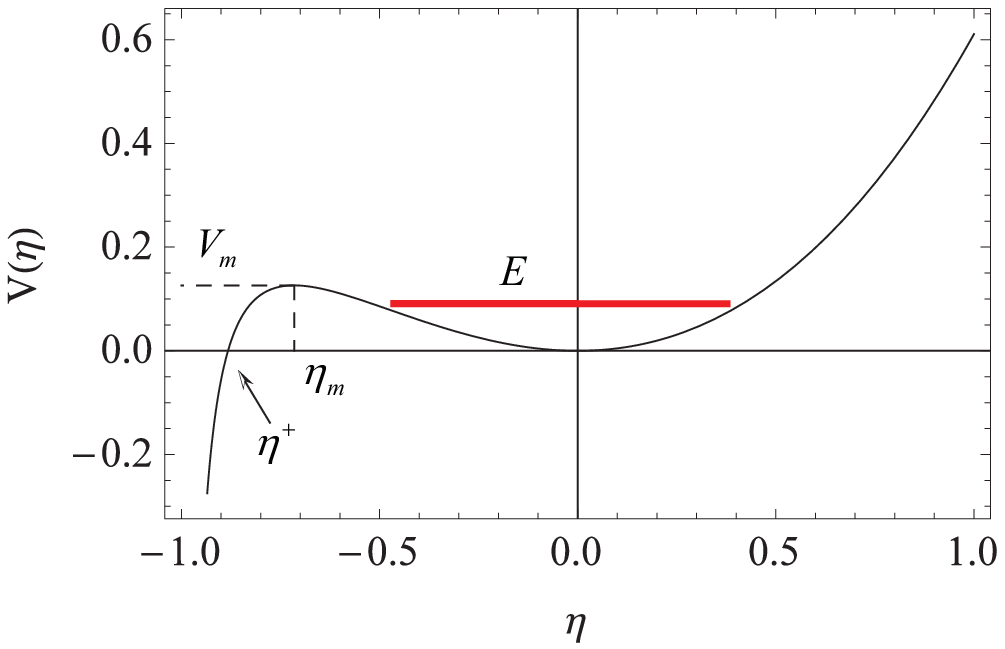} 
\par\end{centering}

\protect\caption{Illustrative plot of $V(\eta)$ verses $\eta$ obtained from Eq.\,(56)
for $M^{2}=0.09$ and $\epsilon_{T}=4/15.$ Here, $\eta^{+}=-0.882,$
$\eta_{m}=-0.715$ and $V(\eta_{m})=0.126.$}

\label{Fig1} 
\end{figure}

Figure \ref{Fig1} shows a schematic plot of $V\left(\eta\right)$
versus $\eta$ for the case where $M^{2}<1$ and $-1<\eta^{+}<0$.
For purpose of illustration, the values of the specific parameters
in Fig.\,\ref{Fig1} have been chosen to be $M^{2}=0.09$ and $\epsilon_{T}=4/15$
in plotting $V\left(\eta\right)$ versus $\eta$. The corresponding
values of $\eta^{+}$, $\eta_{m}$ and $V\left(\eta_{m}\right)$ are
$\eta^{+}=-0.882$, $\eta_{m}=-0.715$, and $V\left(\eta_{m}\right)=0.126$.
For different choices of values for $\epsilon_{T}$ and $M^{2}<1$,
the shape of the $V\left(\eta\right)$ versus $\eta$ curve is qualitatively
similar to that shown in Fig.\,\ref{Fig1}. Referring to Fig.\,\ref{Fig1},
when the effective energy $E$ (the red horizonal line in Fig.\,\ref{Fig1}
lies in the interval $0<E<V\left(\eta_{m}\right)$, Eq.\,(\ref{eq:66})
supports nonlinear periodic solutions for $\eta\left(Z'\right)$ that
oscillate as a function of $Z'$. Here $V\left(\eta_{m}\right)$ is
the local maximum of $V\left(\eta\right)$, which occurs at $\eta=\eta_{m}$
in Fig.\,\ref{Fig1}. Depending on system parameters, these nonlinear
traveling-wave solutions can have large amplitude, representing a
siginificant modulation in beam line density.

\begin{figure}[ptb]
\begin{centering}
\includegraphics[width=3.5in]{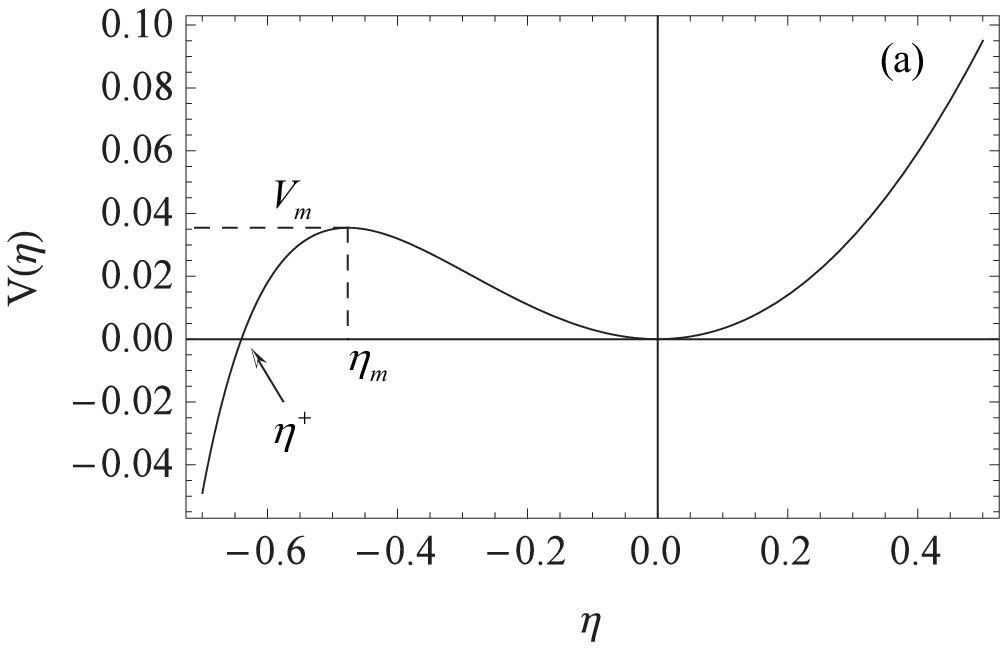} 
\par\end{centering}

\begin{centering}
\includegraphics[width=3.5in]{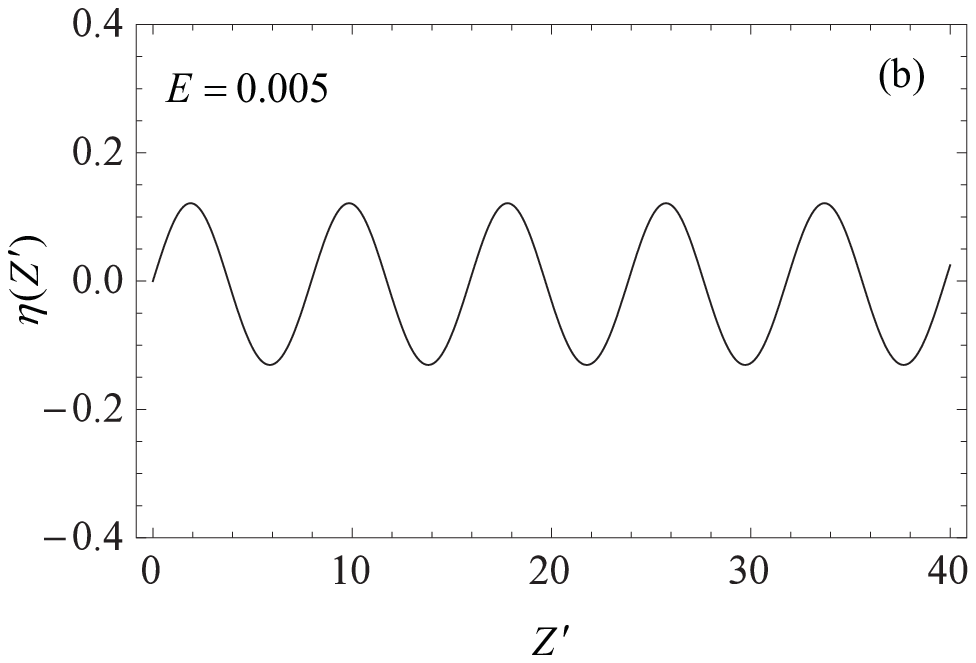}
\par\end{centering}

\begin{centering}
\includegraphics[width=3.5in]{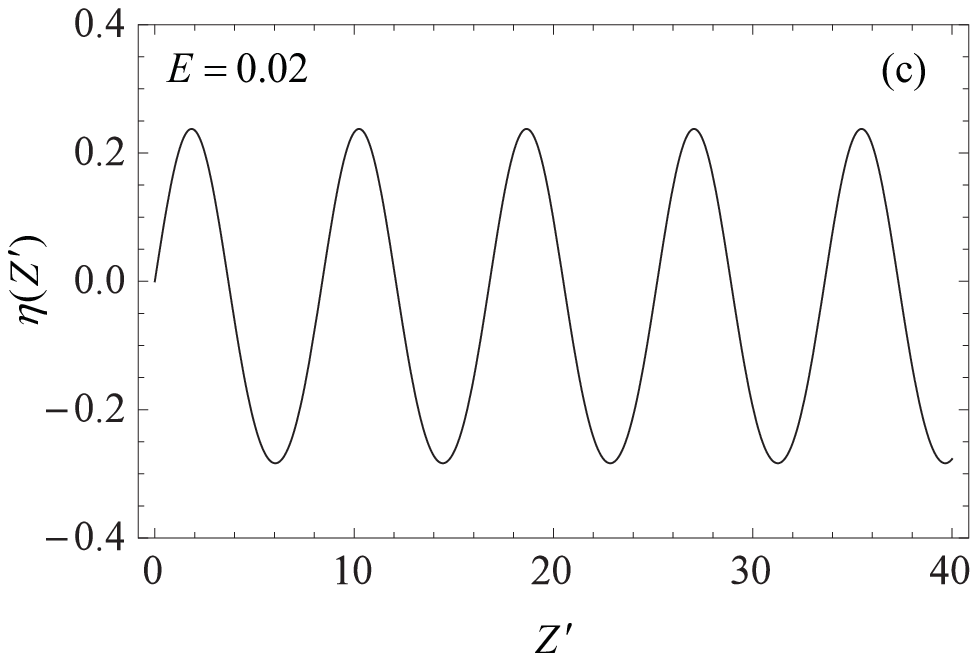} 
\par\end{centering}

\protect\caption{For $M^{2}=0.36,$ $\epsilon_{T}=0,$ $\eta^{+}=-0.64,$ $\eta_{m}=-0.476$
and $V(\eta_{m})=0.0355,$ plots are shown for (a) $V(\eta)$ verses
$\eta;$ (b) $\eta(Z^{\prime})$ verses $Z^{\prime}$ for $\eta'(0)=0.1$
and $E=(1/2)[\eta'(0)]^{2}=0.005;$ and (c) $\eta(Z^{\prime})$ verses
$Z^{\prime}$ for $\eta'(0)=0.2$ and $E=(1/2)[\eta'(0)]^{2}=0.02.$}

\label{Fig2} 
\end{figure}

\begin{figure}[ptb]
\begin{centering}
\includegraphics[width=3.5in]{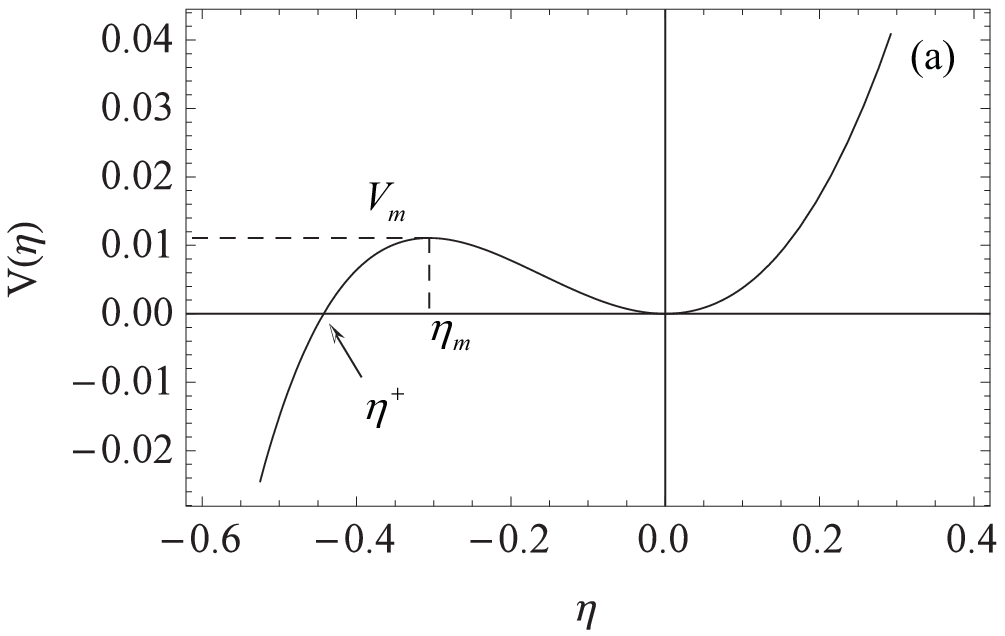} 
\par\end{centering}

\begin{centering}
\includegraphics[width=3.5in]{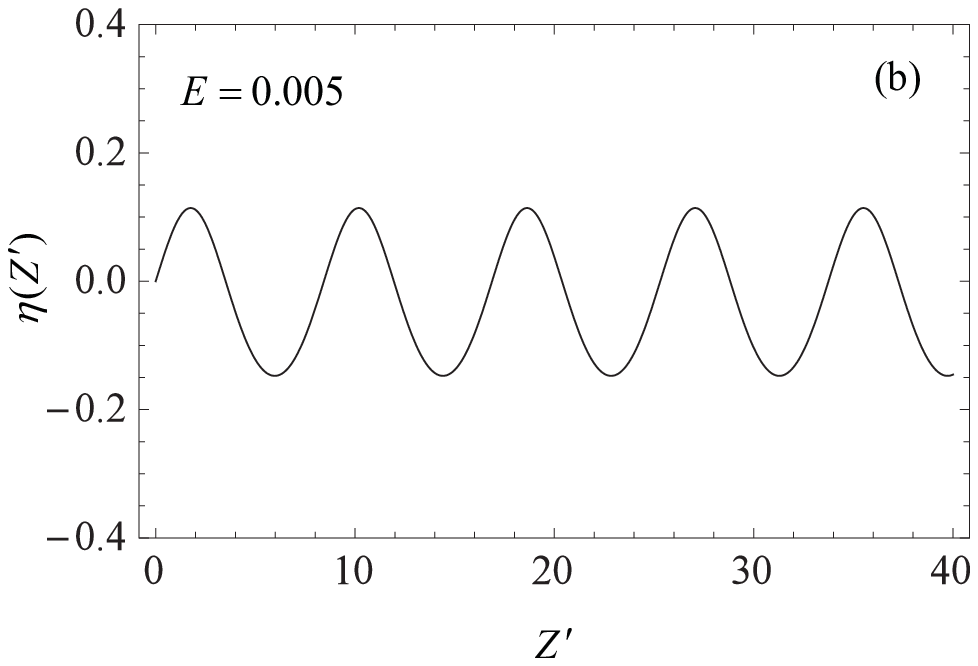}
\par\end{centering}

\begin{centering}
\includegraphics[width=3.5in]{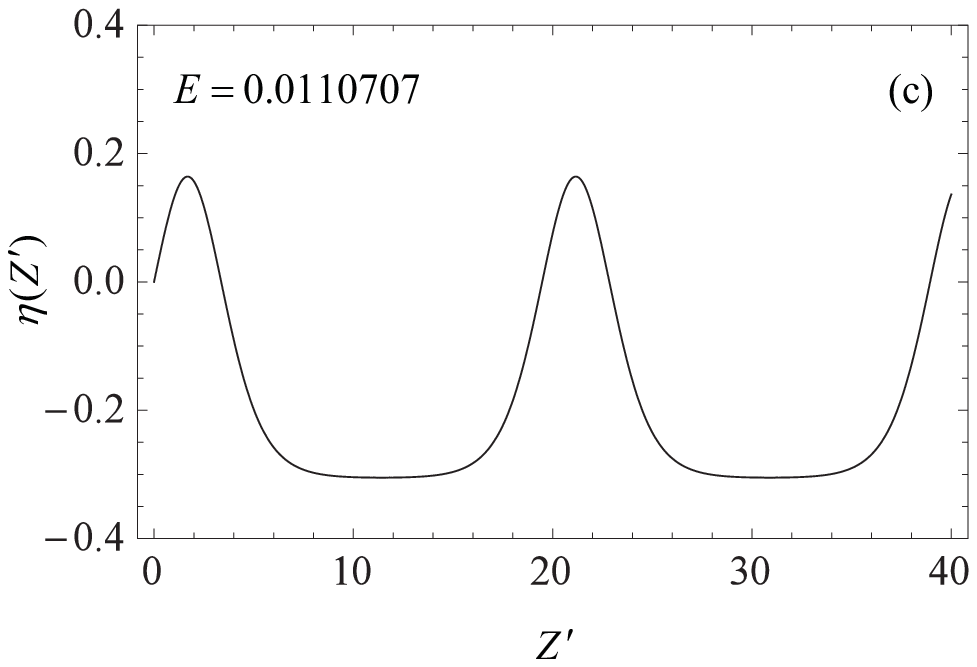} 
\par\end{centering}

\protect\caption{For $M^{2}=0.36,$ $\epsilon_{T}=0.8$, $\eta^{+}=-0.442,$ $\eta_{m}=-0.306$
and $V(\eta_{m})=0.0111,$ plots are shown for (a) $V(\eta)$ verses
$\eta;$ (b) $\eta(Z^{\prime})$ verses $Z^{\prime}$ for $\eta'(0)=0.1$
and $E=(1/2)[\eta'(0)]^{2}=0.005;$ and (c) $\eta(Z^{\prime})$ verses
$Z^{\prime}$ for $\eta'(0)=0.1488$ and $E=(1/2)[\eta'(0)]^{2}=0.0110707.$}

\label{Fig3} 
\end{figure}

\begin{figure}[ptb]
\begin{centering}
\includegraphics[width=3.5in]{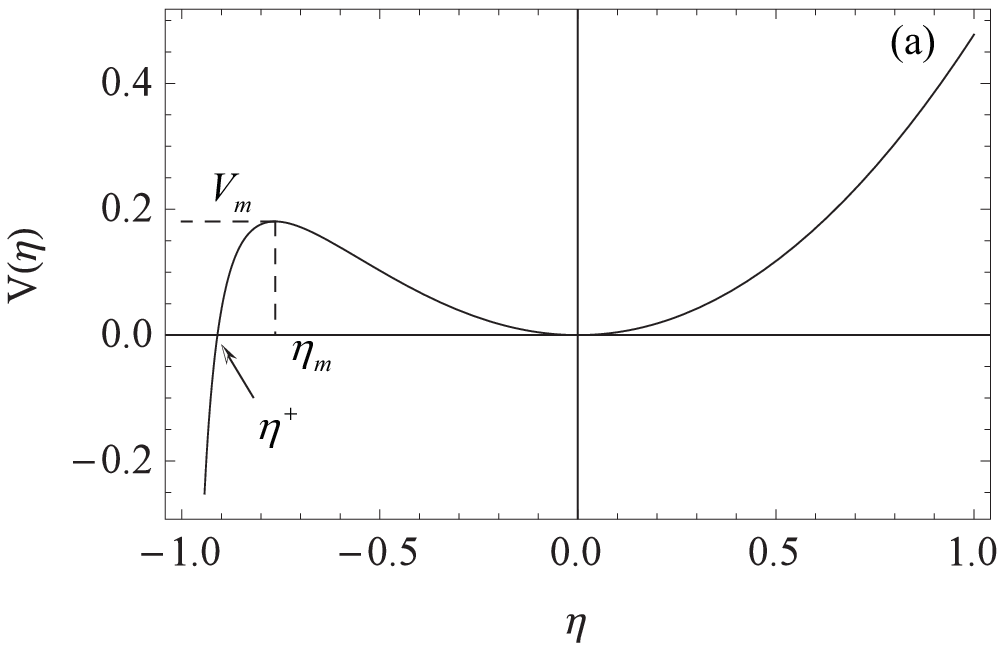} 
\par\end{centering}

\begin{centering}
\includegraphics[width=3.5in]{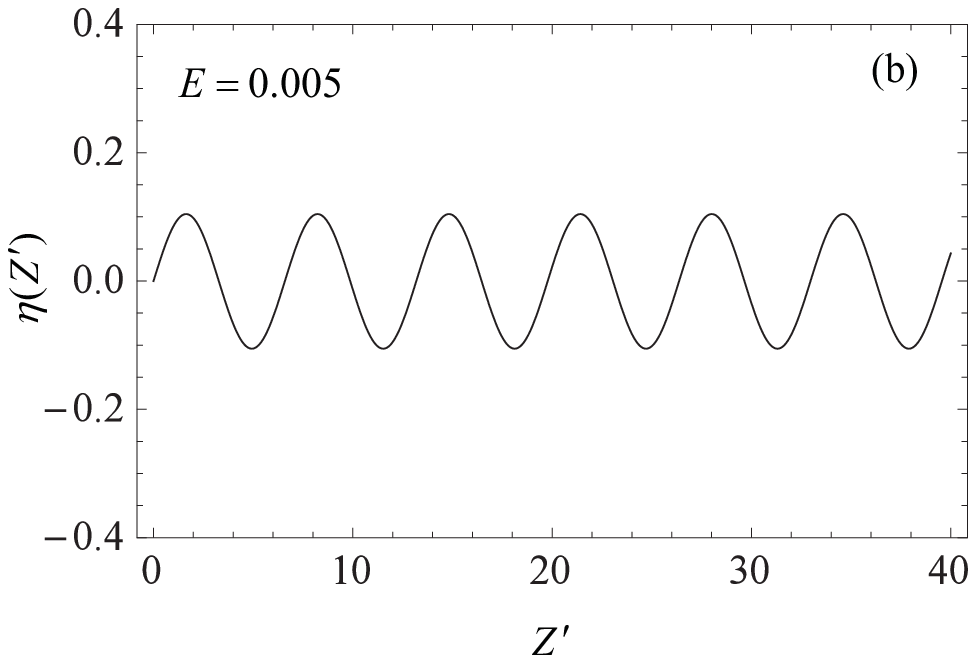}
\par\end{centering}

\begin{centering}
\includegraphics[width=3.5in]{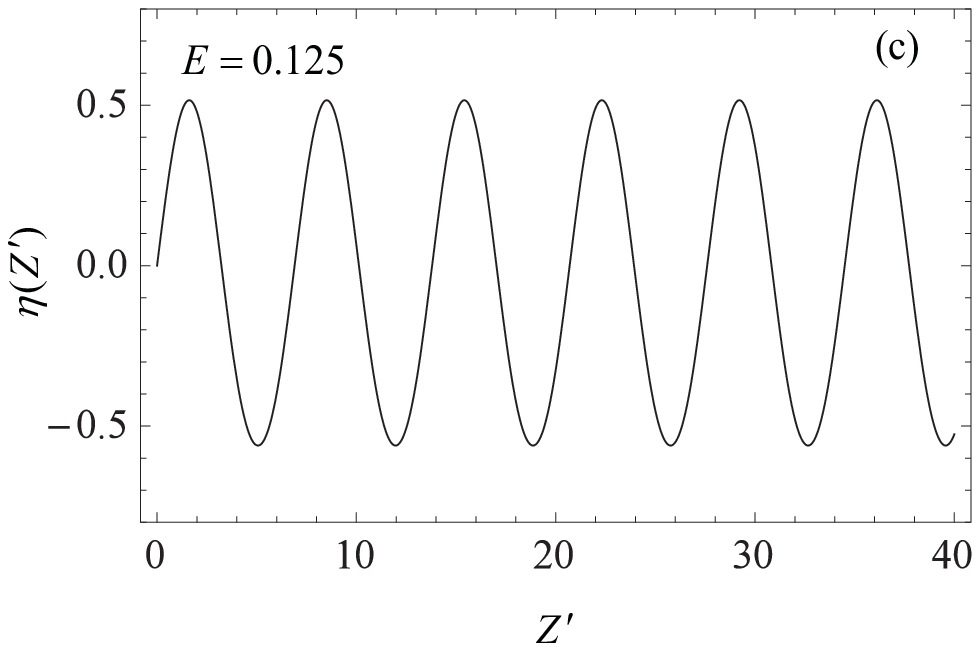} 
\par\end{centering}

\protect\caption{For $M^{2}=0.09,$ $\epsilon_{T}=0$, $\eta^{+}=-0.91,$ $\eta_{m}=-0.764$
and $V(\eta_{m})=0.181,$ plots are shown for (a) $V(\eta)$ verses
$\eta;$ (b) $\eta(Z^{\prime})$ verses $Z^{\prime}$ for $\eta'(0)=0.1$
and $E=(1/2)[\eta'(0)]^{2}=0.005;$ and (c) $\eta(Z^{\prime})$ verses
$Z^{\prime}$ for $\eta'(0)=0.5$ and $E=(1/2)[\eta'(0)]^{2}=0.125.$}

\label{Fig4} 
\end{figure}

\begin{figure}[ptb]
\begin{centering}
\includegraphics[width=3.5in]{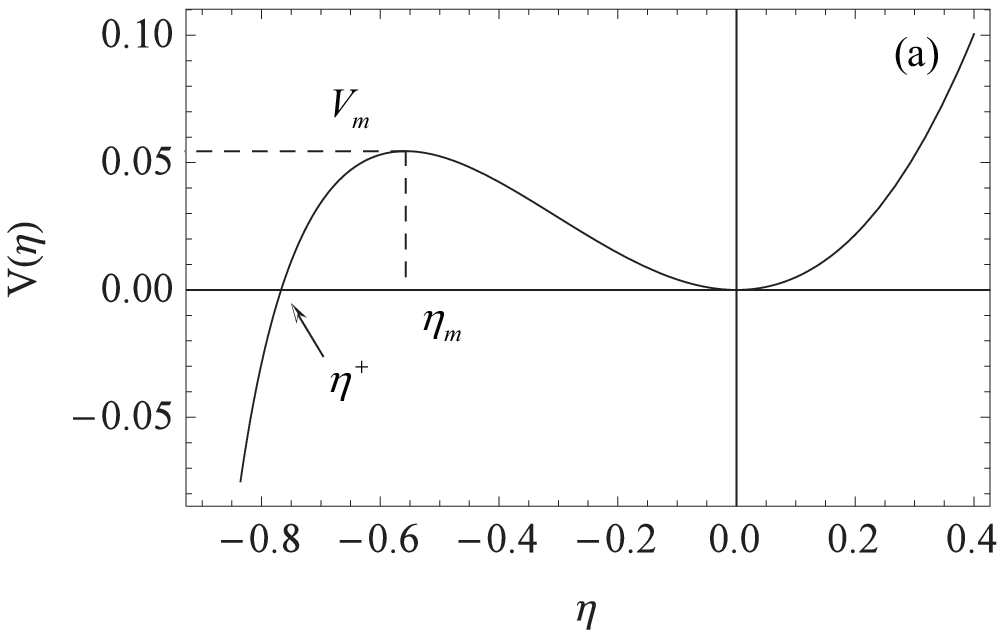} 
\par\end{centering}

\begin{centering}
\includegraphics[width=3.5in]{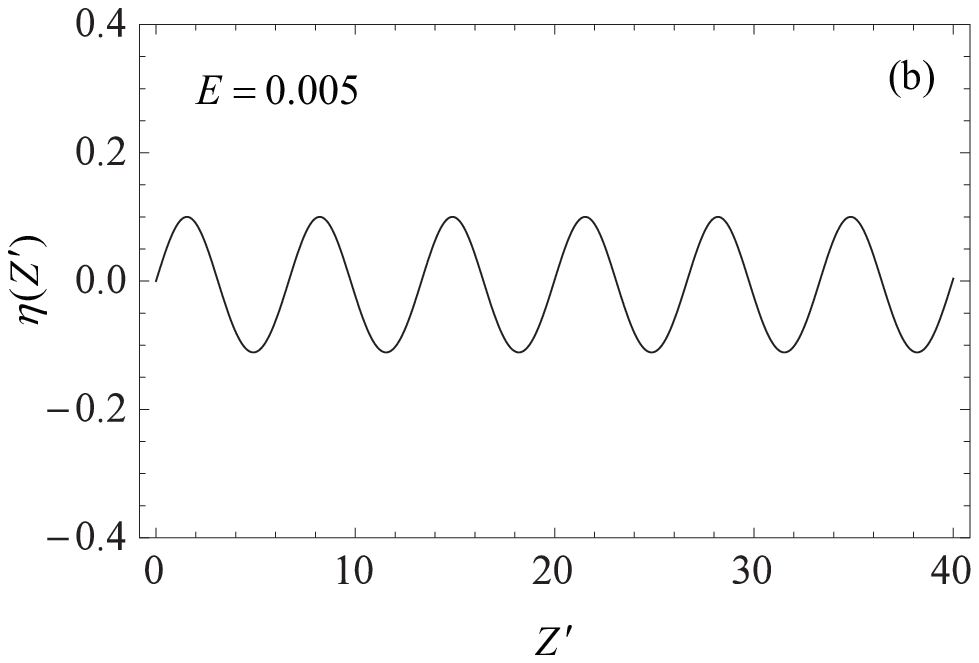}
\par\end{centering}

\begin{centering}
\includegraphics[width=3.5in]{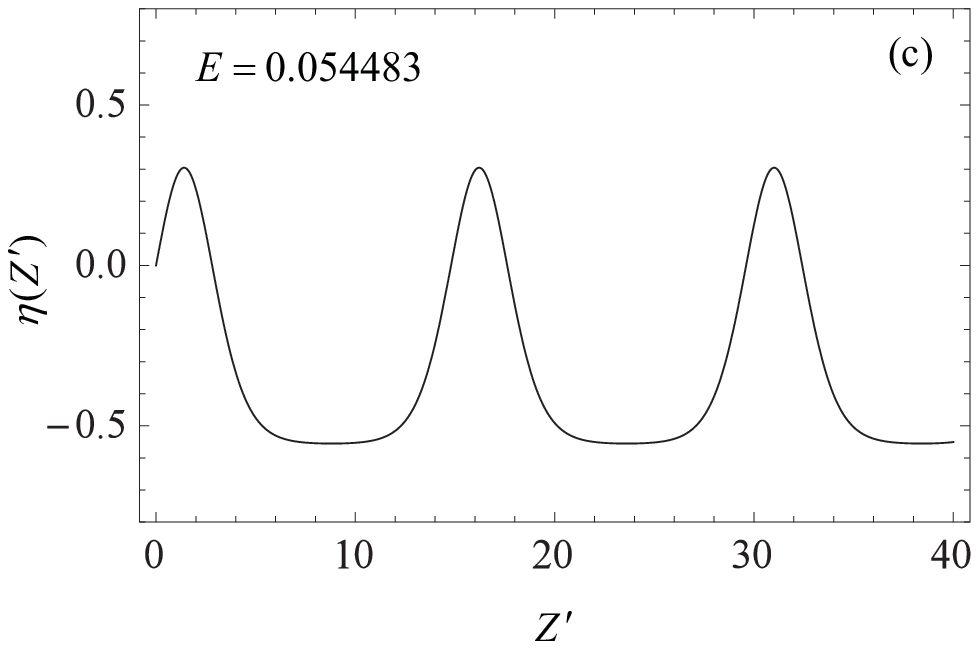} 
\par\end{centering}

\protect\caption{For $M^{2}=0.09,$ $\epsilon_{T}=0.8$, $\eta^{+}=-0.767,$ $\eta_{m}=-0.557$
and $V(\eta_{m})=0.0545,$ plots are shown for (a) $V(\eta)$ verses
$\eta;$ (b) $\eta(Z^{\prime})$ verses $Z^{\prime}$ for $\eta'(0)=0.1$
and $E=(1/2)[\eta'(0)]^{2}=0.005;$ and (c) $\eta(Z^{\prime})$ verses
$Z^{\prime}$ for $\eta'(0)=0.3301$ and $E=(1/2)[\eta'(0)]^{2}=0.054483.$}

\label{Fig5} 
\end{figure}

Referring to the discussion preceding Eq.\,(\ref{eq:58}), the boundary
conditions used to derive Eqs.\,(\ref{eq:55}) and (\ref{eq:56})
from Eqs.\,(\ref{eq:50}) and (\ref{eq:51}) correspond to 

\begin{equation}
\eta\left(0\right)=0=\eta''\left(0\right)\label{eq:63}
\end{equation}
for the class of nonlinear periodic wave solutions obtained from Eq.\,(\ref{eq:62})
when $M^{2}<1$ and $-1<\eta^{+}<0$. Furthermore, from Fig.\,\ref{Fig1}
and Eq.\,(\ref{eq:62}), we note that $V\left(\eta=0\right)=0$ and
the effective energy $E$ can be expressed as

\begin{equation}
E=\frac{1}{2}\left[\eta'\left(0\right)\right]^{2}\,.\label{eq:64}
\end{equation}
Typical numerical solutions for $\eta\left(Z'\right)$, obtained by
integrating Eq.\,(\ref{eq:62}) with $V\left(\eta\right)$ specified
by Eq.\,(\ref{eq:58}), are illustrated in Figs.\,\ref{Fig2}-\ref{Fig5}
for several values of $M^{2}<1$ and $\epsilon_{T}$, and different
values of effective energy level $E$. These correspond to: $M^{2}=0.36$,
$\epsilon_{T}=0$, $E=0.005$ and $E=0.0110707$ (Fig.\,\ref{Fig2});
$M^{2}=0.36$, $\epsilon_{T}=4/15$, $E=0.005$ and $E=0.020$ (Fig.\,\ref{Fig3});
$M^{2}=0.09$, $\epsilon_{T}=0$, $E=0.005$ and $E=0.054483$ (Fig.\,\ref{Fig4});
$M^{2}=0.09$, $\epsilon_{T}=4/15$, $E=0.005$ and $E=0.125$ (Fig.\,\ref{Fig5}).
Close examination of Figs.\,\ref{Fig2}-\ref{Fig5} shows several
interesting trends. First, for smaller values of $M^{2}$, the potential
wells are deeper and broader (compare Figs.\,\ref{Fig2}a and \ref{Fig4}a,
and Figs.\,\ref{Fig3}a and \ref{Fig5}a); and for smaller values
of $\epsilon_{T}$, the potential wells are deeper (compare Figs.\,\ref{Fig2}a
and \ref{Fig3}a, and Figs.\,\ref{Fig4}a and \ref{Fig5}a). Furthermore,
the nonlinear wave amplitude tends to be larger for smaller values
of $M^{2}$ (compare Figs.\,\ref{Fig2}a and \ref{Fig3}a, and Figs.\,\ref{Fig4}a
and \ref{Fig5}a), whereas the wavelength dependance on $M^{2}$ and
$\epsilon_{T}$ tends to be relatively weak (compare Figs.\,\ref{Fig2},
\ref{Fig3}, \ref{Fig4}, \ref{Fig5}). In any case, for $M^{2}<1$,
it is clear from Figs.\,\ref{Fig1}-\ref{Fig5} that Eqs.\,(\ref{eq:62})
and (\ref{eq:58}) support a broad class of nonlinear traveling-wave
solutions for the theoretical model developed here, based on the 1D
kinetic waterbag model for intense beam propagation. Indeed, the modulation
of the beam line density is about $\pm50\%$ for the system parameter
in Figs.\,\ref{Fig4}c and \ref{Fig5}c.

\paragraph{Nonlinear Traveling-pulse (Soliton) Solutions\,$\left(M^{2}>1\ and\ \eta^{+}>0\right)$:}

\begin{figure}[ptb]
\begin{centering}
\includegraphics[width=3.5in]{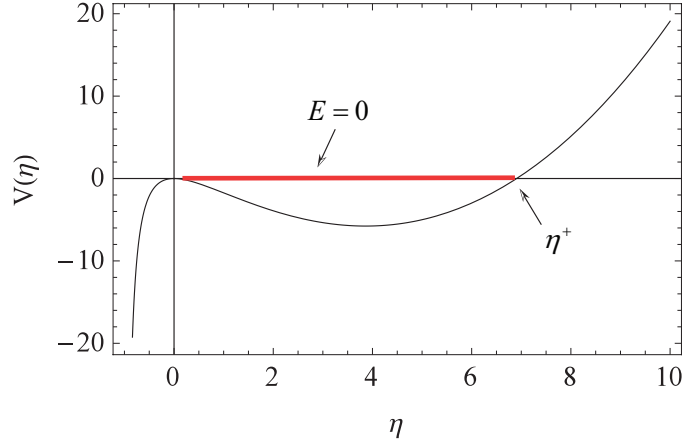} 
\par\end{centering}

\protect\caption{Illustrative plot of $V(\eta)$ verses $\eta$ obtained from Eq.\,(56)
for $M^{2}=9$ and $\epsilon_{T}=1/50.$ Here, $\eta^{+}=6.908,$
and the energy level $E=0$ corresponds to soliton solutions with
maximum amplitude $\eta^{+}=6.908.$ }

\label{Fig6} 
\end{figure}

\begin{figure}[ptb]
\begin{centering}
\includegraphics[width=3.5in]{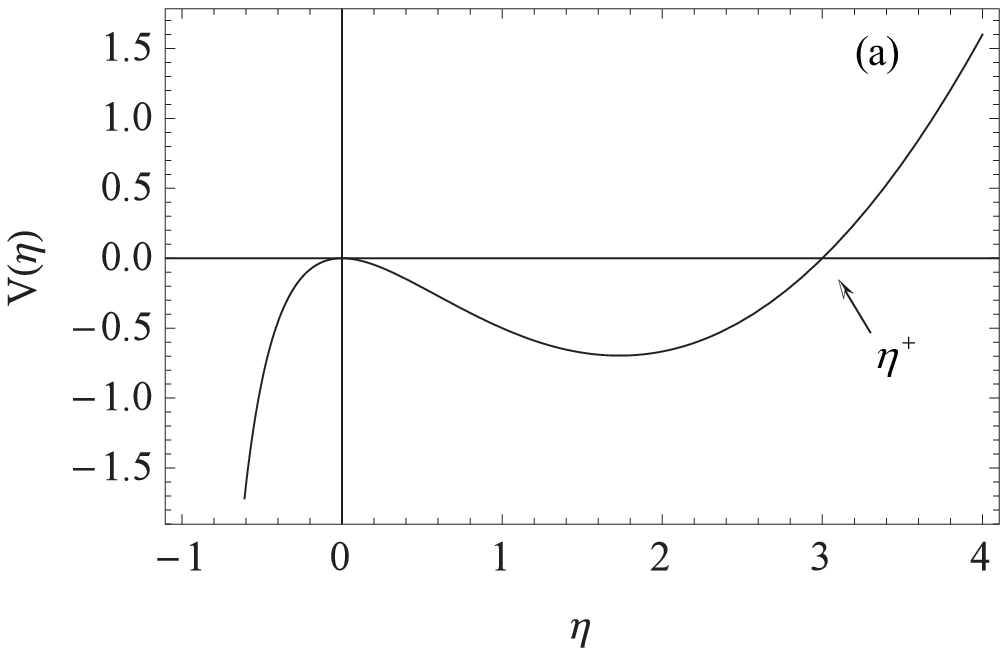} 
\par\end{centering}

\begin{centering}
\includegraphics[width=3.5in]{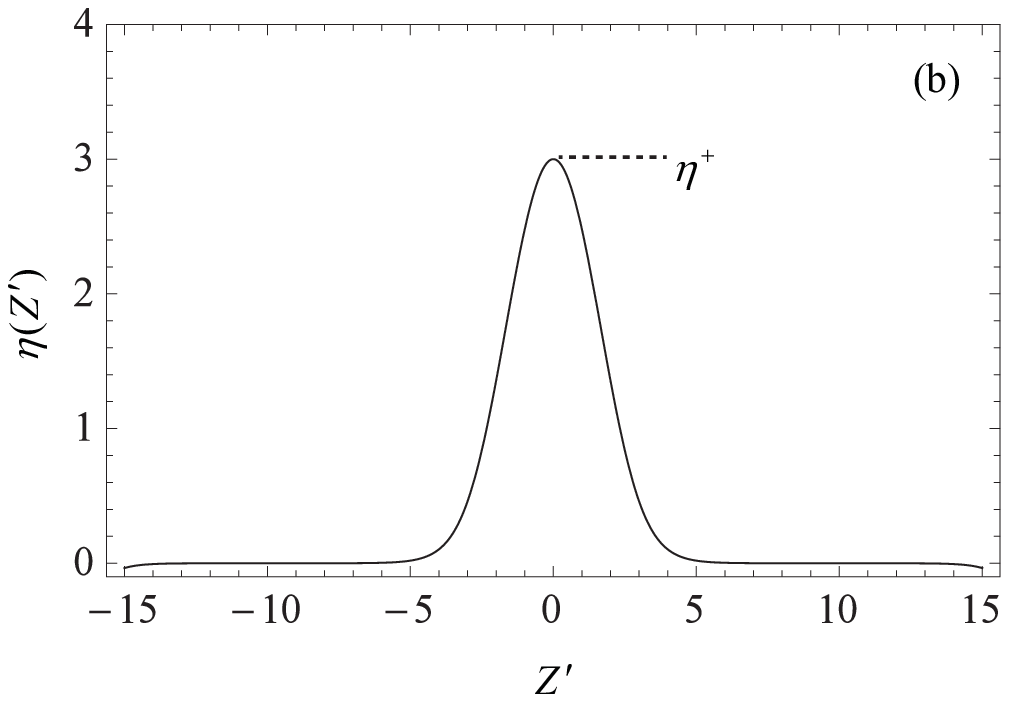}
\par\end{centering}

\protect\caption{Plots of (a) $V(\eta)$ verses $\eta;$ and (b) $\eta(Z^{\prime})$
verses $Z^{\prime}$, obtained from Eqs.\,(56) and (62) for $M^{2}=4,$
$\epsilon_{T}=0$ and $E=0,$ corresponding to soliton amplitudes
$\eta^{+}=3.0.$ }

\label{Fig7} 
\end{figure}

\begin{figure}[ptb]
\begin{centering}
\includegraphics[width=3.5in]{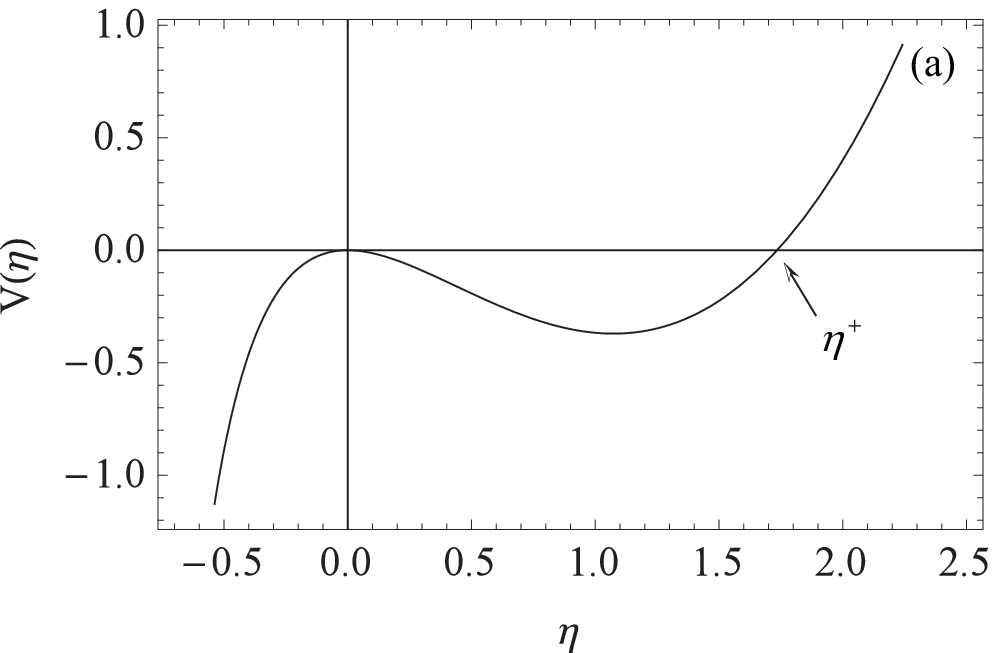} 
\par\end{centering}

\begin{centering}
\includegraphics[width=3.5in]{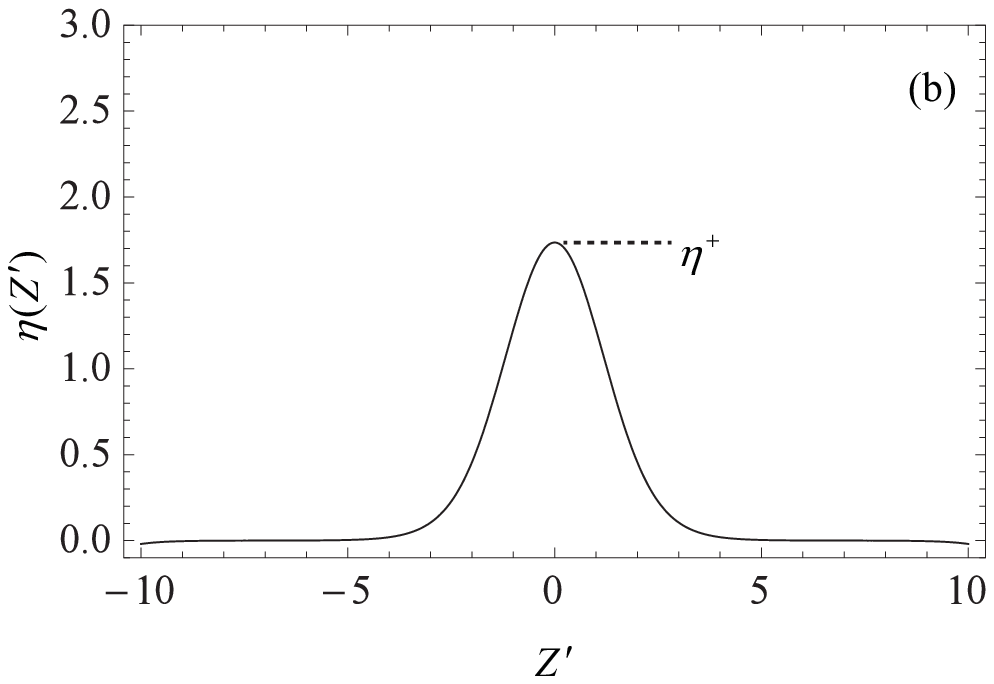}
\par\end{centering}

\protect\caption{Plots of (a) $V(\eta)$ verses $\eta;$ and (b) $\eta(Z^{\prime})$
verses $Z^{\prime}$, obtained from Eqs.\,(56) and (62) for $M^{2}=4,$
$\epsilon_{T}=4/15$ and $E=0,$ corresponding to soliton amplitudes
$\eta^{+}=1.735.$ }

\label{Fig8} 
\end{figure}

\begin{figure}[ptb]
\begin{centering}
\includegraphics[width=3.5in]{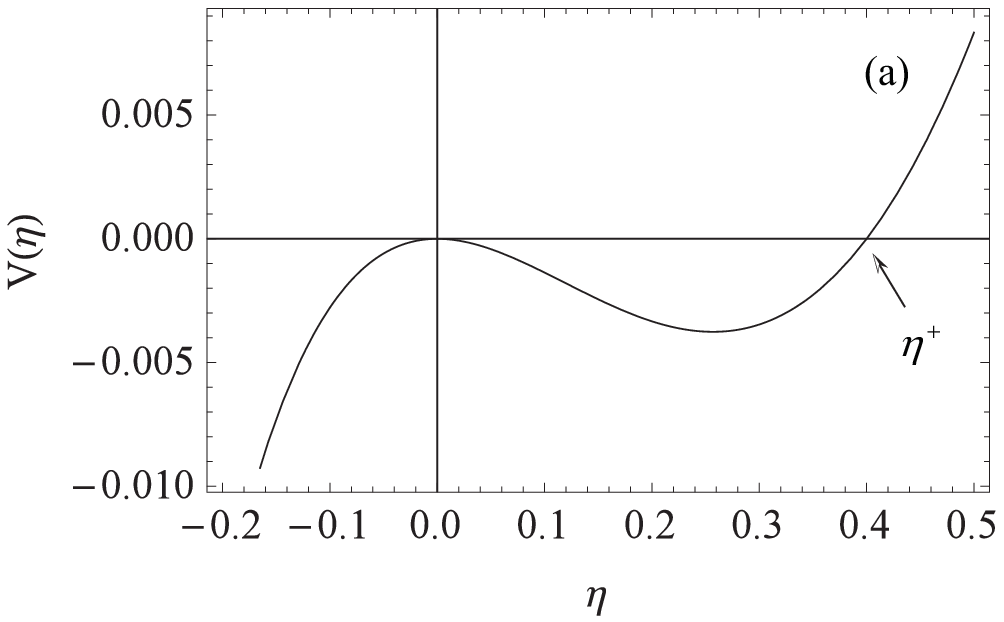} 
\par\end{centering}

\begin{centering}
\includegraphics[width=3.5in]{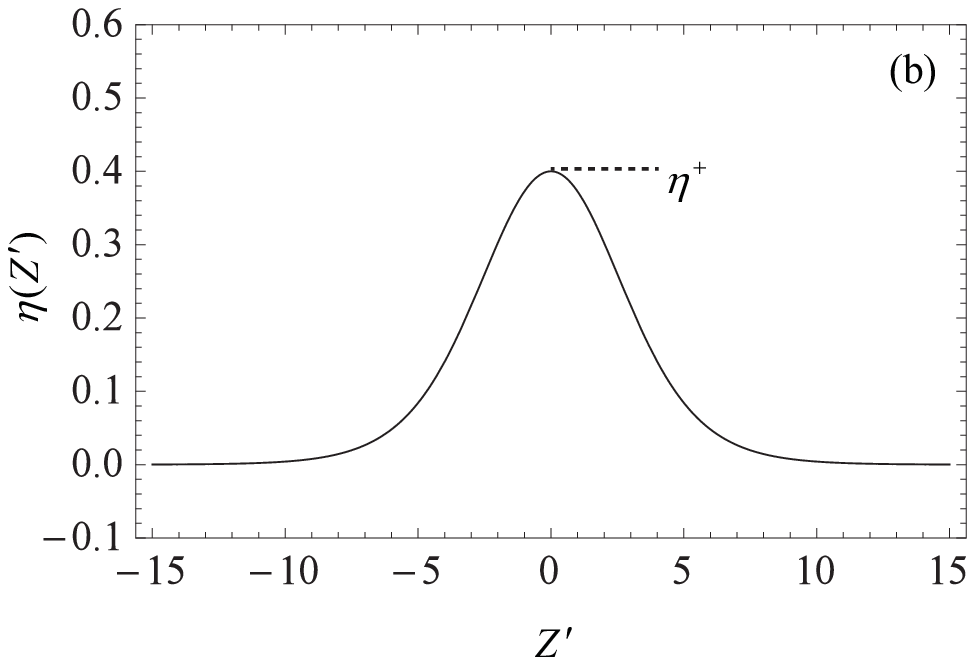}
\par\end{centering}

\protect\caption{Plots of (a) $V(\eta)$ verses $\eta;$ and (b) $\eta(Z^{\prime})$
verses $Z^{\prime}$, obtained from Eqs.\,(56) and (62) for $M^{2}=1.2,$
$\epsilon_{T}=0$ and $E=0,$ corresponding to soliton amplitudes
$\eta^{+}=0.4.$ }

\label{Fig9} 
\end{figure}

\begin{figure}[ptb]
\begin{centering}
\includegraphics[width=3.5in]{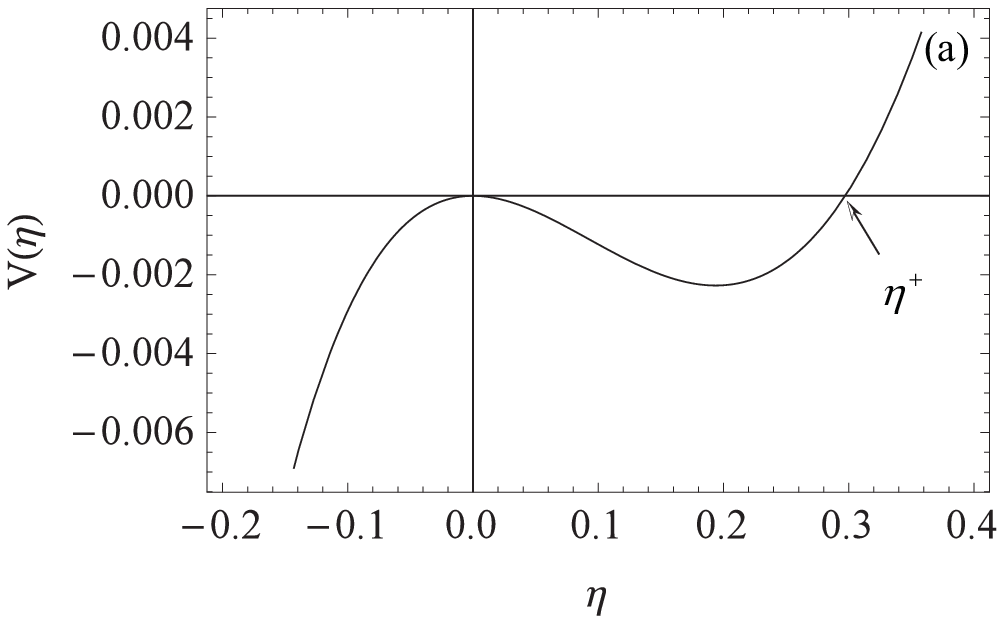} 
\par\end{centering}

\begin{centering}
\includegraphics[width=3.5in]{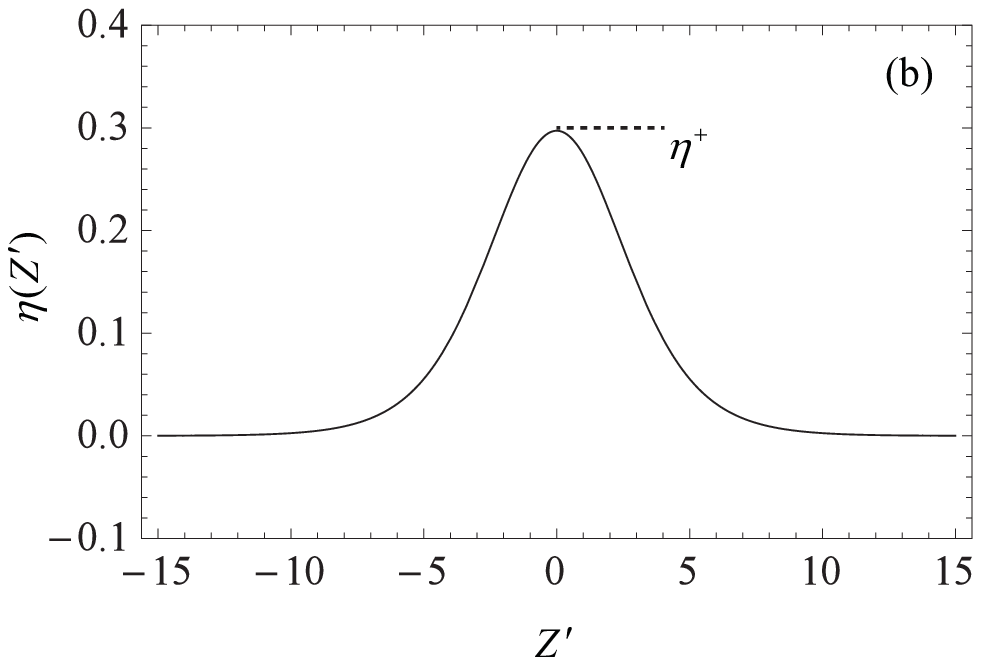}
\par\end{centering}

\protect\caption{Plots of (a) $V(\eta)$ verses $\eta;$ and (b) $\eta(Z^{\prime})$
verses $Z^{\prime}$, obtained from Eqs.\,(56) and (62) for $M^{2}=1.2,$
$\epsilon_{T}=4/15$ and $E=0,$ corresponding to soliton amplitudes
$\eta^{+}=0.297.$ }

\label{Fig10} 
\end{figure}

We now consider Eqs.\,(\ref{eq:62}) and Eq.\,(\ref{eq:58}) {[}
or equivalently, Eq.\,(\ref{eq:56}){]} in circumstences where $M^{2}>1$
and $\eta^{+}>0$. In this case, the effective potential has the qualitative
shape illustrated in Fig.\,\ref{Fig6}, which has been plotted for
the choice of parameters $M^{2}=9$ and $\epsilon_{T}=1/30$. The
physically allowed, localized pulse solutions (soliton solutions)
corresponds to the energy level

\begin{equation}
E=0\,,\label{eq:65}
\end{equation}
which is the red horizontal line in Fig.\,\ref{Fig6}, and boundary
conditions

\begin{equation}
\eta\left(Z'=\pm\infty\right)=0=\eta''\left(Z'=\pm\infty\right)\label{eq:66}
\end{equation}
discussed prior to Eq.\,(\ref{eq:58}). Referring to Fig.\,\ref{Fig6},
when Eq.\,(\ref{eq:62}) is integrated forward from $Z'=-\infty$
where $\eta=0$, the perturbed line density, $\eta$ increases monotonically
through positive values to a maximum amplitude $\eta^{+}$ (the soliton
amplitude) and then decreases monotonically to $\eta=0$ when $Z'=+\infty$.
The regime where $M^{2}>1$ by a sufficiently large amount corresponds
to a strongly nonlinear regime where the density compression is large
with $\eta^{+}>1$. On the other hand, when $M^{2}-1=\epsilon$ is
small with $0<\epsilon\ll1$, the soliton amplitude is correspondingly
small. This will become apparent from the numerical solutions to Eqs.\,(\ref{eq:62})
and (\ref{eq:58}) consistent with Eqs.\,(\ref{eq:65}) and (\ref{eq:66})
presented later in this section in Figs.\,\ref{Fig7}-\ref{Fig10}.

Typical numerical solutions to Eqs.\,(\ref{eq:62}) and (\ref{eq:58}),
subject to Eqs.\,(\ref{eq:65}) and (\ref{eq:66}), are illustrated
in Figs.\,\ref{Fig7}-\ref{Fig10} for several values of $M^{2}>1$
and $\epsilon_{T}$. These correspond to: $M^{2}=4$ and $\epsilon_{T}=0$
(Fig.\,\ref{Fig7}); $M^{2}=4$ and $\epsilon_{T}=4/15$ (Fig.\,\ref{Fig8});
$M^{2}=1.2$ and $\epsilon_{T}=0$ (Fig.\,\ref{Fig9}); and $M^{2}=1.2$
and $\epsilon_{T}=4/15$ (Fig.\,\ref{Fig10}). Close examination
of Figs.\,\ref{Fig7}-\ref{Fig10} shows that the soliton amplitude
increases with increasing $M^{2}$ (compare Figs.\,\ref{Fig7} and
\ref{Fig8} with Figs.\,\ref{Fig9} and \ref{Fig10}), reaching a
highly nonlinear regime with $\eta^{+}=3.0$ in Fig.\,\ref{Fig7}
and $\eta^{+}=1.735$ in Fig.\,\ref{Fig8}, where $M^{2}=4$. In
contrast, the soliton width tends show a relatively weak dependence
on longitudinal velocity spread, as measured by $\epsilon_{T}$ (compare
Fig.\,\ref{Fig8} with Fig.\,\ref{Fig7}, and Fig.\,\ref{Fig10}
with Fig.\,\ref{Fig9}). It's clear from Figs.\,\ref{Fig7}-\ref{Fig10}
that the soliton solutions to Eqs.\,(\ref{eq:62}) and (\ref{eq:58})
exhibit a strong nonlinear dependence on $M^{2}$, and can correspond
to highly compressed line density for sufficiently large $M^{2}$.

In the special circumstances where $M^{2}$ exceeds $1$ by a small
amount, i.e.\,, $M^{2}=1+\Delta$ where $0<\Delta\ll1$. it is readily
shown that Eq.\,(\ref{eq:54}) can be approximated for small $\eta$
by 

\begin{equation}
\frac{\partial^{2}\eta}{\partial Z'^{2}}+\left\{ \left[\frac{3}{2}M^{2}+\frac{3}{2}\epsilon_{T}\right]\eta-\left(M^{2}-1\right)\right\} \eta=0\label{eq:67}
\end{equation}
Equation (\ref{eq:67}) can be solved exactly for $\eta\left(Z'\right)=\lambda_{b}\left(Z'\right)/\lambda_{b0}-1$
to give

\begin{equation}
\eta\left(Z'\right)=\left(\frac{M^{2}-1}{M^{2}+\epsilon_{T}}\right)\mathrm{{sech}^{2}}\left[\frac{1}{2}\left(M^{2}-1\right)^{1/2}\left(Z-MT\right)\right]\,.\label{eq:68}
\end{equation}
Note that the soliton amplitude in Eq.\,(\ref{eq:68}) is small for
$M^{2}=1+\Delta$ with $\Delta\ll1$. Also, the $\mathrm{sech^{2}\left\{ \cdots\right\} }$
pulse shape in Eq.\,(\ref{eq:68}) is similar to the soliton pulse
shape obtained from the Korleweq-deVries equation in the weakly nonlinear
regime \cite{15davidson2004korteweg}.

Finally, it should be noted that the oscillatory solutions obtained
from Eqs.\,(\ref{eq:58}) and (\ref{eq:62}) when $M^{2}>1$ and
the energy level $E$ in Fig.\,\ref{Fig6} is negative with $V_{min}<E<0$
are not considered here. These solutions are unphysical because they
oscillate about a positive non-zero average value of $\bar{\eta}=\bar{\lambda}_{b}/\lambda_{b0}-1>0$,
rather than oscillate about $\bar{\eta}=\bar{\lambda}_{b}/\lambda_{b0}-1\approx0$,
as occurs in Figs.\,\ref{Fig2}-\ref{Fig5} when $M^{2}<1$.

\section{Coherent nonlinear structures obtained from fully kinetic g-factor
model\label{sec:Kinetic}}

The kinetic waterbag model developed in Sec.\,\ref{sec:Waterbag}
of this paper has clearly demonstrated the rich variety of coherent
nonlinear structures supported by the 1D kinetic model based on Eqs.\,(\ref{eq:9})
and (\ref{eq:10}) {[}or equivalently, Eqs.\,(\ref{eq:1}) and (\ref{eq:2}){]}
for the specific choice of waterbag distribution $F_{b}\left(z,p_{z},t\right)$
in Eq.\,(\ref{eq:31}). In this Section, we examine solutions to
Eqs.\,(\ref{eq:9}) and (\ref{eq:10}) for an even broader class
of distribution functions $F_{b}\left(z,p_{z},t\right)$, recognizing
that Eqs.\,(\ref{eq:9}) and (\ref{eq:10}) are Galilean invariant.
That is, if we transform variables to a frames of reference moving
with constant longitudinal velocity $V_{0}=const.$ according to $z'=z-V_{0}t$,
$p_{z}'=p_{z}-m_{b}V_{0}$, $t'=t$, then in the new dynamical variables
$\left(z',p_{z}',t'\right)$, the equations for $F_{b}\left(z',p_{z}',t'\right)$
and $\left\langle \phi\right\rangle \left(z',t'\right)$ are identical
in form to Eqs.\,(\ref{eq:9}) and (\ref{eq:10}). Time-stationary
solutions ($\partial/\partial t'=0$) in the new variables $\left(z',p_{z}',t'\right)$
then correspond to undistorted traveling-wave or traveling-pulse solutions
moving with constant velocity $V_{0}=const.$ in the original variables
$\left(z,p_{z},t\right)$. The present analysis of Eqs.\,(\ref{eq:9})
and (\ref{eq:10}) parallels the original Bernstein-Greene-Kruskal
(BGK) formulation of BGK solutions to the 1D Vlasov-Poisson equations
\cite{25bernstein1957exact,26SeeMethod2}, except for the fact that
Eq.\,(\ref{eq:10}), which connects $\left\langle \phi\right\rangle \left(z,t\right)$
to the line density $\lambda_{b}\left(z,t\right)$, has a very different
structure than the 1D Poisson equation.

Referring to Eqs.\,(\ref{eq:9}) and (\ref{eq:10}), we introduce
the scaled dimensionless variables ($Z,P_{z},T$) defined by

\begin{align}
Z=\left(\frac{U_{b0}^{2}+U_{bT}^{2}}{U_{b2}^{2}}\right)^{\frac{1}{2}}\frac{z}{r_{w}}\,, & \thinspace\thinspace\thinspace T=\left(\frac{U_{b0}^{2}+U_{bT}^{2}}{U_{b2}^{2}}\right)\frac{U_{b2}t}{r_{w}}\,,\nonumber \\
P_{z}=\frac{p_{z}}{m_{b}\left(U_{b0}^{2}+U_{bT}^{2}\right)^{1/2}} & =\frac{v_{z}}{\left(U_{b0}^{2}+U_{bT}^{2}\right)^{1/2}}\equiv V_{z}\,,\label{eq:69}
\end{align}
where $U_{b0}^{2}$ and $U_{b2}^{2}$ are defined in Eq.\,(\ref{eq:3}),
and $U_{bT}^{2}=const.$ is the longitudinal velocity spread characteristic
of the distribution function $F_{b}$. We futher introduce the dimensionless
distribution function $\hat{F}_{b}\left(Z,P_{z},T\right)$ defined
by 

\begin{equation}
\hat{F}_{b}=\lambda_{b0}^{-1}\frac{F_{b}}{m_{b}\left(U_{b0}^{2}+U_{bT}^{2}\right)^{1/2}}\,,\label{eq:70}
\end{equation}
where $\lambda_{b0}=const.$ is the characteristic line density of
the beam particles, e.g., the average value. From Eqs.\,(\ref{eq:69}),
(\ref{eq:70}) and the definition of line density $\lambda_{b}=\int dp_{z}\,F_{b}$,
it follows that the perturbation in line density $\eta=\lambda_{b}/\lambda_{b0}-1$
can be expressed as

\begin{equation}
\eta\left(Z,T\right)=\int dP_{z}\,\hat{F}_{b}\left(Z,P_{z},T\right)-1\,,\label{eq:71}
\end{equation}
where the $P_{z}$ integration covers the range $-\infty<P_{z}<\infty$
in Eq.\,(\ref{eq:71}). Transforming variables according to Eqs.\,(\ref{eq:69})
and (\ref{eq:70}), and making use of Eq.\,(\ref{eq:71}), it is
readily shown that Eqs.\,(\ref{eq:9}) and (\ref{eq:10}) can be
expressed in the new variables as

\begin{equation}
\frac{\partial\hat{F}_{b}}{\partial T}+V_{z}\frac{\partial\hat{F}_{b}}{\partial Z}-\frac{\partial\psi}{\partial Z}\frac{\partial\hat{F}_{b}}{\partial P_{z}}=0\,,\label{eq:72}
\end{equation}
and

\begin{equation}
\psi=\eta+\frac{\partial^{2}\eta}{\partial Z^{2}}\,,\label{eq:73}
\end{equation}
where $\psi\left(Z,T\right)$ is the normalized (dimensionless) potential
defined by

\begin{equation}
\psi=\frac{e\left\langle \phi\right\rangle }{m_{b}\left(U_{b0}^{2}+U_{bT}^{2}\right)^{1/2}}\,,\label{eq:74}
\end{equation}
Equations (\ref{eq:72}) and (\ref{eq:73}), where $\eta$ and $\int dP_{z}\,\hat{F}_{b}$
are related by Eq.\,(\ref{eq:71}), constitute coupled nonlinear
equations describing the self-consistent evolution of the distribution
function $\hat{F}_{b}\left(Z,P_{z},T\right)$, normalized potential
$\psi\left(Z,T\right)$, and normalized perturbed line density $\eta\left(Z,T\right)$.
Equations (\ref{eq:71})-(\ref{eq:73}) are fully equivalent to the
original dynamical equations (\ref{eq:9})-(\ref{eq:11}), and can
be used to investigate 1D kinetic properties of the nonlinear beam
dynamics over a wide range of system parameters.

Keeping in mind that Eqs.\,(\ref{eq:72}) and (\ref{eq:73}) are
Galilean invariant, if we transform Eqs.\,(\ref{eq:72}) and (\ref{eq:73})
from the variables $\left(Z,P_{z},T\right)$ to a frame moving with
normalized velocity $M=const.$ according to $Z'=Z-MT$, $V_{z}'=V_{z}-M$,
$T'=T$, then Eqs.\,(\ref{eq:72}) and (\ref{eq:73}) have exactly
the same form in the new variables, with $\left(Z,P_{z},T\right)$
replaced by $\left(Z',P_{z}',T'\right)$, i.e.,

\begin{equation}
\frac{\partial\hat{F}_{b}}{\partial T'}+V_{z}'\frac{\partial\hat{F}_{b}}{\partial Z'}-\frac{\partial\psi}{\partial Z'}\frac{\partial\hat{F}_{b}}{\partial P_{z}'}=0\,,\label{eq:75}
\end{equation}
and

\begin{equation}
\psi=\eta+\frac{\partial^{2}\eta}{\partial Z'^{2}}\,.\label{eq:76}
\end{equation}
Here, $P_{z}'=V_{z}'$, and $\hat{F}_{b}\left(Z',P_{z}',T'\right)$
and $\eta\left(Z',P_{z}',T'\right)$ are related by

\begin{equation}
\eta\left(Z',T'\right)=\int dP_{z}'\,F_{b}\left(Z',P_{z}',T'\right)-1\,.\label{eq:77}
\end{equation}
Therefore, the traveling-pulse or traveling-wave solutions that have
stationary profile shape in the primed variables $\left(Z',P_{z}',T'\right)$
are determined by setting $\partial/\partial T'=0$ in Eqs.\,(\ref{eq:75})-(\ref{eq:77}).

Setting $\partial\hat{F}_{b}/\partial T'=0$ in Eq.\,(\ref{eq:75})
gives for $\hat{F}_{b}\left(Z',P_{z}'\right)$

\begin{equation}
V_{z}'\frac{\partial\hat{F}_{b}}{\partial Z'}-\frac{\partial\psi}{\partial Z'}\frac{\partial\hat{F}_{b}}{\partial P_{z}'}=0\,,\label{eq:78}
\end{equation}
where $\psi\left(Z'\right)$ and $\eta\left(Z'\right)$ solve Eq.\,(\ref{eq:76}),
and $\eta\left(Z'\right)$ is related to $\hat{F}_{b}\left(Z',P_{z}'\right)$
by Eq.\,(\ref{eq:77}). We introduce the energy variable $W'$ defined
by

\begin{equation}
W'=\frac{1}{2}V_{z}'^{2}+\psi\left(Z'\right)\,.\label{eq:79}
\end{equation}
Then the solution to Eq.\,(\ref{eq:78}) for $\hat{F}_{b}\left(Z',P_{z}'\right)$
can be expressed as

\begin{equation}
\hat{F}_{b}\left(Z',V_{z}'\right)=\hat{F}_{b}^{>}\left(W'\right)\Theta\left(V_{z}'\right)+\hat{F}_{b}^{<}\left(W'\right)\Theta\left(-V_{z}'\right)\,,\label{eq:80}
\end{equation}
where

\begin{equation}
\Theta\left(V_{z}'\right)=\begin{cases}
1,\, & for\ V_{z}'>0\,,\\
0\,, & for\ V_{z}'<0\,.
\end{cases}\label{eq:81}
\end{equation}
Note from Eq.\,(\ref{eq:79}) that

\begin{equation}
dV_{z}'=\pm dW'/\left[2\left(W'-\psi\right)\right]^{1/2}\,,\label{eq:82}
\end{equation}
where $+$ corresponds to $V_{z}'>0$, and $-$ corresponds to $V_{z}'<0$.
Substituting Eqs.\,(\ref{eq:80}) and (\ref{eq:81}) into Eq.\,(\ref{eq:77})
gives

\begin{equation}
\eta=\int_{\psi}^{\infty}dW'\,\frac{\left[\hat{F}_{b}^{>}\left(W'\right)+\hat{F}_{b}^{<}\left(W'\right)\right]}{\left[2\left(W'-\psi\right)\right]^{1/2}}-1\,,\label{eq:83}
\end{equation}
which relate the perturbation in beam line density $\eta\left(Z'\right)$
to the potential $\psi\left(Z'\right)$ and the distribution functions
$\hat{F}_{b}^{>}\left(W'\right)$ and $\hat{F}_{b}^{<}\left(W'\right)$.

\begin{figure}[ptb]
\begin{centering}
\includegraphics[width=3.5in]{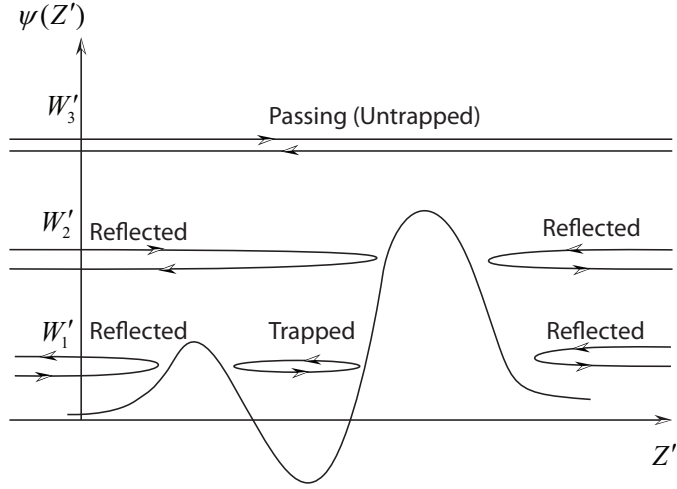} 
\par\end{centering}

\protect\caption{Illustrative plot of the effective potential $\psi(Z^{\prime})$ verses
$Z^{\prime}$ occurring in Eq.\,(33) showing the three classes of
particle orbits corresponding to (a) passing (untrapped) particles
with energy $W_{3}^{\prime}$, (b) reflected particles with energy
$W_{2}^{\prime}$, and (c) reflected or trapped particles (depending
on the range of $Z^{\prime}$) with energy $W_{1}^{\prime}$. The
form of $\psi(Z^{\prime})$ in Fig. 11 corresponds to an isolated
pulse with $\psi(Z^{\prime}\rightarrow\pm\infty)=0.$ }

\label{Fig11} 
\end{figure}

Figure \ref{Fig11} shows an illustrative plot of the potential $\psi\left(Z'\right)$
as a function of $Z'$. Depending on the values of the energy $W'$
and range of $Z'$, there are three classes of particle orbits: (a)
particles that are reflected from the potential; (b) particles that
are trapped and undergo periodic motion; and (c) passing (untrapped)
particles that don't change direction, but pass over the potential
maximum, first slowing down and then speeding up during the motion.
For the trapped particles and the reflected particles, it follows
that $\hat{F}^{>}\left(W'\right)=\hat{F}^{<}\left(W'\right)$ so that

\begin{equation}
\hat{F}_{Tr}\left(W'\right)=\hat{F}_{Tr}^{<}\left(W'\right)+\hat{F}_{Tr}^{>}\left(W'\right)=2\hat{F}_{Tr}^{<}\left(W'\right)=2\hat{F}_{Tr}^{>}\left(W'\right)\label{eq:84}
\end{equation}
and

\begin{equation}
\hat{F}_{Ref}\left(W'\right)=\hat{F}_{Ref}^{<}\left(W'\right)+\hat{F}_{Ref}^{>}\left(W'\right)=2\hat{F}_{Ref}^{<}\left(W'\right)=2\hat{F}_{Ref}^{>}\left(W'\right)\,.\label{eq:85}
\end{equation}
On the other hand, for the passing (untrapped) particles, $\hat{F}_{Un}^{>}\left(W'\right)$
and $\hat{F}_{Un}^{<}\left(W'\right)$ can be specified independently,
depending on whether the particles have $V_{z}'>0$ or $V_{z}'<0$,
respectively.

\begin{figure}[ptb]
\begin{centering}
\includegraphics[width=3.5in]{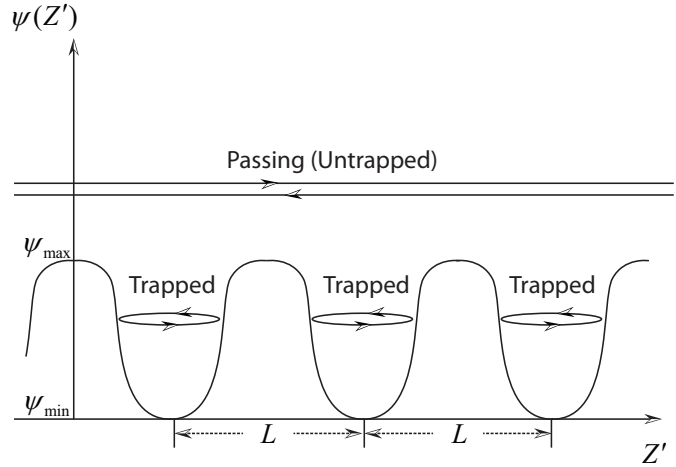} 
\par\end{centering}

\protect\caption{Illustrative plot of the effective potential $\psi(Z^{\prime})$ verses
$Z^{\prime}$ for the case where $\psi(Z^{\prime})$ has a nonlinear
periodic waveform with $\psi(Z^{\prime}+L)=\psi(Z^{\prime}),$ where
$L$ is the periodicity length. In the figure, passing particles with
energy $W'>\psi_{max}$ are untrapped, whereas particles with energy
$\psi_{min}<W'<\psi_{max}$ are trapped and exhibit periodic motion
in the potential $\psi(Z^{\prime})$.}

\label{Fig12} 
\end{figure}

The form of $\psi\left(Z'\right)$ shown in Fig.\,\ref{Fig11} corresponds
to a stationary isolated pulse in primed variables, with $\psi\left(Z'=\pm\infty\right)=0$.
By contrast, Fig.\,\ref{Fig12} shows a plot of $\psi\left(Z'\right)$
versus $Z'$ for the case where $\psi\left(Z'\right)$ has a periodic
nonlinear wave structure with

\begin{equation}
\psi\left(Z'+L\right)=\psi\left(Z'\right).\label{eq:86}
\end{equation}
From Fig.\,\ref{Fig12}, trapped particles with energy $W'$ in the
range

\begin{equation}
\psi_{min}<W'<\psi_{max}\label{eq:87}
\end{equation}
exhibit periodic motion. On the other hand, passing particles with
energy $W'$ in the range (see Fig.\,\ref{Fig12})

\begin{equation}
W'>\psi_{max}\label{eq:88}
\end{equation}
correspond to untrapped particles that pass over the potential $\psi\left(Z'\right)$,
periodically speeding up and slowing down, but not changing their
direction of motion. Furthermore, for the nonlinear periodic waveform
for the potential $\psi\left(Z'+L\right)=\psi\left(Z'\right)$ shown
in Fig.\,\ref{Fig12}, it follows from Eqs.\,(\ref{eq:76}) and
(\ref{eq:83}) that the waveform for the perturbation in line charge
also satisfies $\eta\left(Z'+L\right)=\eta\left(Z'\right)$. Here,
$\eta\left(Z'\right)$ is related to $\psi\left(Z'\right)$ and the
trapped-particle and untrapped-particle distribution functions by
Eq.\,(\ref{eq:83}), which gives

\begin{equation}
1+\eta=\int_{\psi}^{\psi_{max}}dW'\,\frac{\hat{F}_{Tr}\left(W'\right)}{\left[2\left(W'-\psi\right)\right]^{1/2}}+\int_{\psi}^{\infty}dW'\,\frac{\hat{F}_{Un}\left(W'\right)}{\left[2\left(W'-\psi\right)\right]^{1/2}}\,.\label{eq:89}
\end{equation}
In Eq.\,(\ref{eq:89}), the integration over the trapped-particle
distributuion $\hat{F}_{Tr}\left(W'\right)$ is over the interval
of $W'$ corresponding to $\psi_{min}<\psi<W'<\psi_{max}$, and the
integration over the untrapped particle distribution $\hat{F}_{Un}\left(W'\right)$
is over the interval of $W'$ corresponding to $\psi_{max}<\psi<W'<\infty$.

Equations (\ref{eq:76}) and (\ref{eq:89}) can be used to determine
detailed properties of self-consistent nonlinear periodic solutions
for $\eta\left(Z'\right)$ and $\psi\left(Z'\right)$ for a broad
range of choices of $\hat{F}_{Tr}\left(W'\right)$ and $\hat{F}_{Un}\left(W'\right)$.
Furthermore, depending on system parameters, the amplitudes of the
wave perturbations can range from small to moderately large amplitude.
For purposes of illustration the procedure for solving Eqs.\,(\ref{eq:76})
and (\ref{eq:89}) for the case of nonlinear periodic solutions for
$\eta\left(Z'\right)$ and $\psi\left(Z'\right)$, we consider the
special case where $\hat{F}_{Tr}\left(W'\right)=0$, and the untrapped
distribution function has the monoenergetic form

\begin{equation}
\hat{F}_{Un}\left(W'\right)=A\sqrt{2W_{U}'}\delta\left(W'-W_{U}'\right)\,,\label{eq:90}
\end{equation}
where $W_{U}'=const.$\,, $A=const.$\,, and $W_{U}'>\psi_{max}$
(see Fig.\,\ref{Fig12}). Substituting $\hat{F}_{Tr}\left(W'\right)=0$
and Eq.\,(\ref{eq:90}) into Eq.\,(\ref{eq:89}) readily gives

\begin{equation}
1+\eta\left(Z'\right)=\frac{A}{\left[1-\psi\left(Z'\right)/W_{U}'\right]^{1/2}}\,.\label{eq:91}
\end{equation}
For present purpose, we choose the normalization constant $A$ in
Eq.\,(\ref{eq:91}) such that the line density perturbation $\eta\left(Z'\right)=\lambda_{b}\left(Z'\right)/\lambda_{b0}-1$
and potential perturbation $\psi\left(Z'\right)$ are simultaneously
zero for all $Z'$, i.e.\,, $\eta\left(Z'\right)=0$ for all $Z'$
when $\psi\left(Z'\right)=0$. From Eq.\,(\ref{eq:91}), this readily
gives $A=1$ for the value of the constant $A$. Squaring Eq.\,(\ref{eq:91})
and solving for $\psi\left(Z'\right)$ when $A=1$ readily gives

\begin{equation}
\psi=W_{U}'\left[1-\frac{1}{\left(1+\eta\right)^{2}}\right]\,.\label{eq:92}
\end{equation}
Note that Eq.\,(\ref{eq:92}) determines $\psi\left(Z'\right)$ as
a function of $\eta\left(Z'\right)$, which can be substituted into
Eq.\,(\ref{eq:76}) to solve for $\eta\left(Z'\right)$.

Similar to the analysis in Sec.\,\ref{sub:Coherent-Nonlinear-Traveling}
for the class of nonlinear periodic traveling-wave solutions with
$\eta\left(Z'+L\right)=\eta\left(Z'\right)$ and $\psi\left(Z'+L\right)=\psi\left(Z'\right)$,
we examine Eqs.\,(\ref{eq:76}) and (\ref{eq:92}) for the case where
the boundary conditions correspond to $\eta\left(Z'=0\right)=0$ and
$\left[\partial^{2}\eta/\partial Z'^{2}\right]_{Z'=0}=0$. Substituting
Eq.\,(\ref{eq:92}) into Eq.\,(\ref{eq:76}) we readily obtain

\begin{equation}
\frac{\partial^{2}\eta}{\partial Z'^{2}}+\eta=W_{U}'\left[1-\frac{1}{\left(1+\eta\right)^{2}}\right]\,,\label{eq:93}
\end{equation}
which can also be expressed as

\begin{equation}
\frac{\partial^{2}\eta}{\partial Z'^{2}}+\frac{\partial V}{\partial\eta}=0\,,\label{eq:94}
\end{equation}
where

\begin{align}
\frac{\partial V}{\partial\eta} & =\eta-W_{U}'\left[1-\frac{1}{\left(1+\eta\right)^{2}}\right]\nonumber \\
 & =\frac{\eta}{\left(1+\eta\right)^{2}}\left[\eta^{2}+\left(2-W_{U}'\right)\eta+\left(1-2W_{U}'\right)\right]\,.\label{eq:95}
\end{align}
Note that Eq.\,(\ref{eq:94}) has the form of a dynamical equation
of motion, with $\eta$ playing the role of displacement, $Z'$ playing
the role of time, and $V\left(\eta\right)$ playing the role of an
effective potential. Making use of Eq.\,(\ref{eq:95}), it is readily
shown that

\begin{equation}
\frac{\partial^{2}V}{\partial\eta{}^{2}}=1-\frac{2W_{U}'}{\left(1+\eta\right)^{3}}\,,\label{eq:96}
\end{equation}
and

\begin{align}
V\left(\eta\right) & =\frac{1}{2}\eta^{2}-W_{U}'\left[\eta+\frac{1}{1+\eta}-1\right]\nonumber \\
 & =\frac{1}{2}\frac{\eta^{2}}{\left(1+\eta\right)^{2}}\left\{ \eta+\left[1-2W_{U}'\right]\right\} \,,\label{eq:97}
\end{align}
where the constant of integration in Eq.\,(\ref{eq:97}) has heen
chosen so that $V\left(\eta=0\right)=0$.

\begin{figure}[ptb]
\begin{centering}
\includegraphics[width=3.5in]{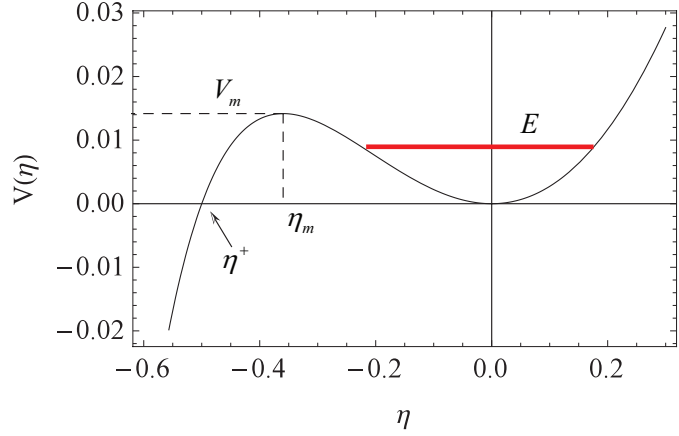} 
\par\end{centering}

\protect\caption{Plot of $V(\eta)$ verses $\eta$ obtained from Eq.\,(97) for $2W_{u}^{\prime}=0.5$
and $\eta^{+}=-0.5.$ Here, $\eta_{m}=-0.36$, and $V(\eta_{m})=0.014.$
Nonlinear periodic solutions for $\eta(Z^{\prime})$ exist for energy
level $E$ in the range $0<E<V(\eta_{m}).$}

\label{Fig13} 
\end{figure}

Close examination of Eqs.\,(\ref{eq:93})-(\ref{eq:96}) shows that
Eq.\,(\ref{eq:94}) supports oscillatory solutions for $\eta\left(Z'\right)$
about $\eta=0$ provided $\left[\partial^{2}V/\partial\eta^{2}\right]_{\eta=0}>0$,
or equivalently, 

\begin{equation}
2W_{U}'<1\,.\label{eq:98}
\end{equation}
When the inequality in Eq.\,(\ref{eq:98}) is satisfied, the plot
of $V\left(\eta\right)$ versus $\eta$ has the characteristic shape
illustrated in Fig.\,\ref{Fig13} for the choice of parameter $2W_{U}'<1$.
Here, $V\left(\eta\right)$ has a minimum at $\eta=0$, and passed
through zero at

\begin{equation}
\eta=\eta^{+}=-\left[1-2W_{U}'\right]\,,\label{eq:99}
\end{equation}
where $V\left(\eta=\eta^{+}\right)=0$ {[}see Eqs.\,(\ref{eq:97})
and (\ref{eq:99}){]}. Similar to the analysis in Sec.\,\ref{sub:Coherent-Nonlinear-Traveling},
Eq.\,(\ref{eq:94}) can be integrated to give the energy conservation
relation $\left(1/2\right)\left[\partial\eta/\partial Z'\right]^{2}+V\left(\eta\right)=E=const.$
{[}see also Eq.\,(\ref{eq:62}){]}, where $E=\left(1/2\right)\left[\eta'\left(0\right)\right]^{2}$
is the effective energy level. Referring to Fig.\,\ref{Fig13}, Eqs.\,(\ref{eq:94})
and (\ref{eq:97}) support nonlinear periodic oscillatory solutions
for $\eta\left(Z'\right)$ for $E$ in the range $0<E<V_{m}$, where
$V_{m}\equiv V\left(\eta=\eta_{m}\right)$ is the local maximum of
$V\left(\eta\right)$ at $\eta=\eta_{m}$. For the choice of dimensionless
parameter $2W_{u}'=1/2$ in Fig.\,\ref{Fig13}, it is readilty shown
that $\eta_{m}=-0.360$ and $V_{m}=V\left(\eta=\eta_{m}\right)=0.014$.

\begin{figure}[ptb]
\begin{centering}
\includegraphics[width=3.5in]{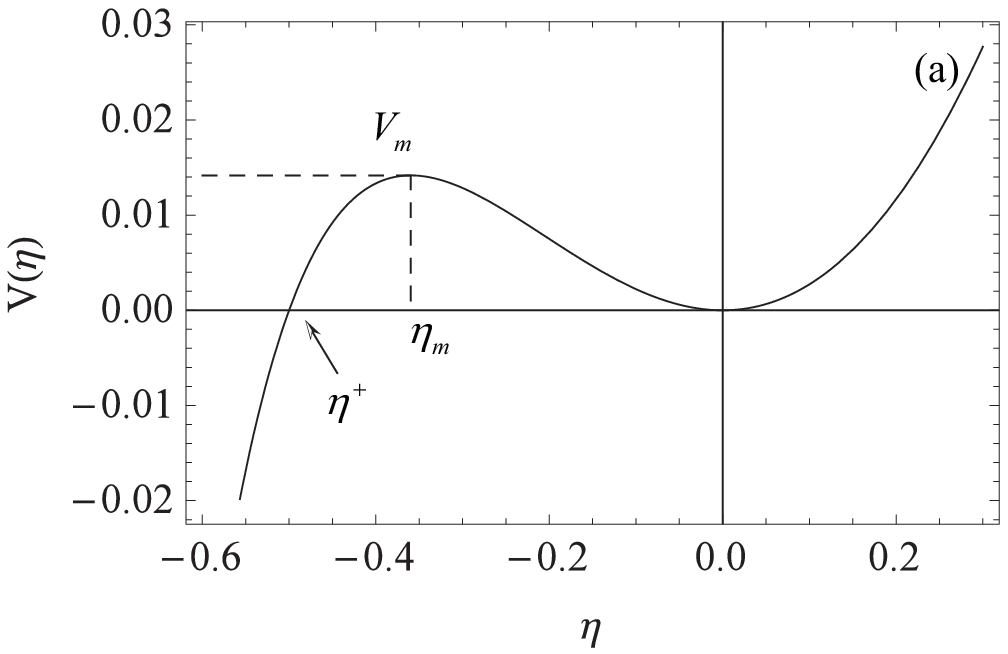} 
\par\end{centering}

\begin{centering}
\includegraphics[width=3.5in]{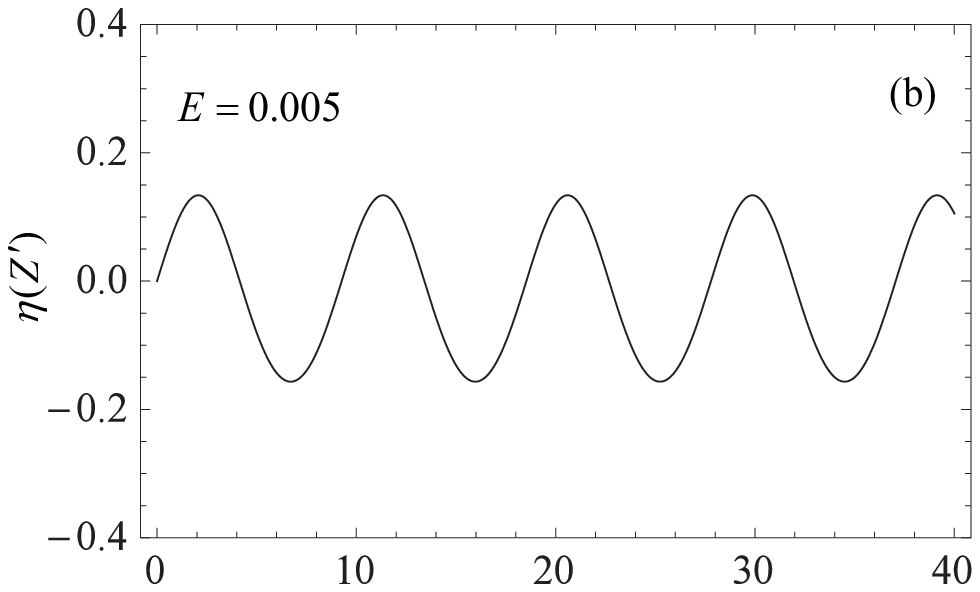}
\par\end{centering}

\protect\caption{Plots are shown for (a) $V(\eta)$ verses $\eta$, and (b) $\eta(Z^{\prime})$
verses $Z^{\prime}$, obtained from Eq.\,(94) for $M^{2}=0.5,$ $\eta'(0)=0.1,$
$E=1/2[\eta'(0)]^{2}=0.005,$ $\eta^{+}=-0.5,$ $\eta_{m}=-0.36$
and $V(\eta_{m})=0.014.$}

\label{Fig14} 
\end{figure}

\begin{figure}[ptb]
\begin{centering}
\includegraphics[width=3.5in]{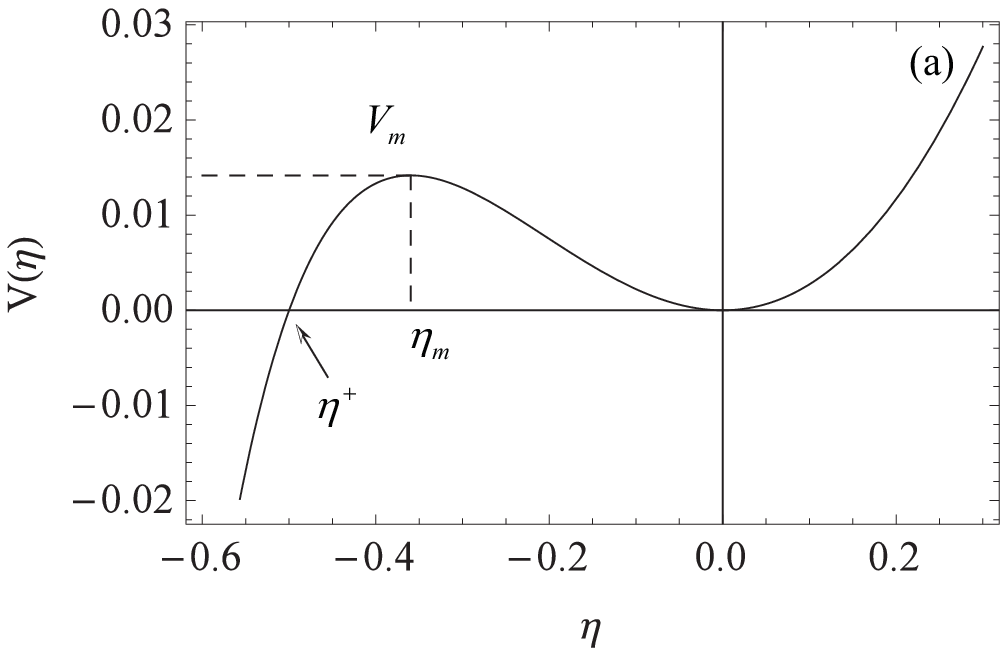} 
\par\end{centering}

\begin{centering}
\includegraphics[width=3.5in]{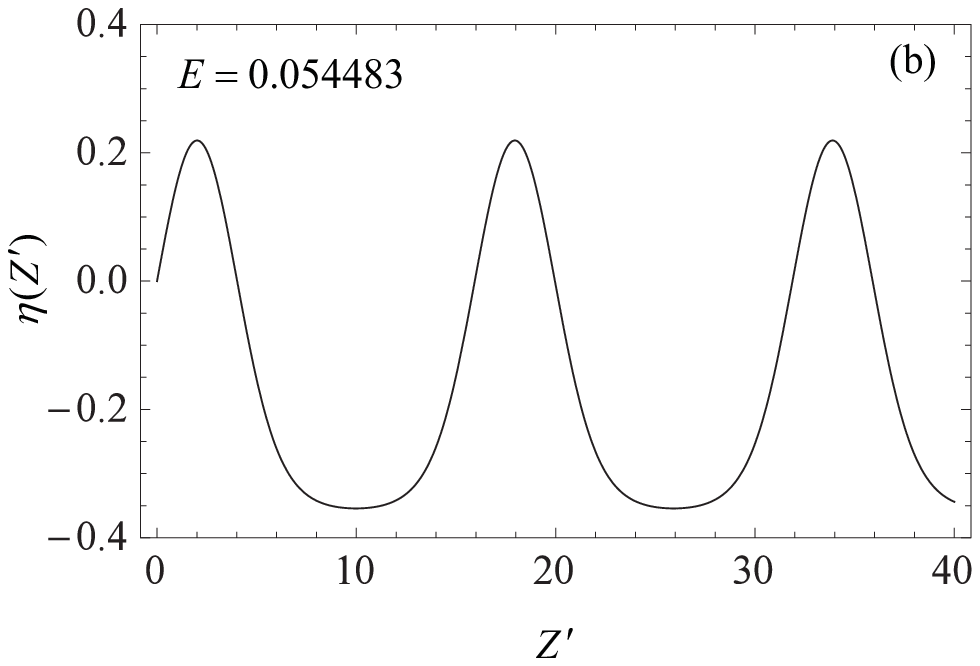}
\par\end{centering}

\protect\caption{Plots are shown for (a) $V(\eta)$ verses $\eta$, and (b) $\eta(Z^{\prime})$
verses $Z^{\prime}$, obtained from Eq.\,(94) for $M^{2}=0.5,$ $\eta'(0)=0.1863,$
$E=1/2[\eta'(0)]^{2}=0.054883,$ $\eta^{+}=-0.5,$ $\eta_{m}=-0.36$
and $V(\eta_{m})=0.014.$}

\label{Fig15} 
\end{figure}

\begin{figure}[ptb]
\begin{centering}
\includegraphics[width=3.5in]{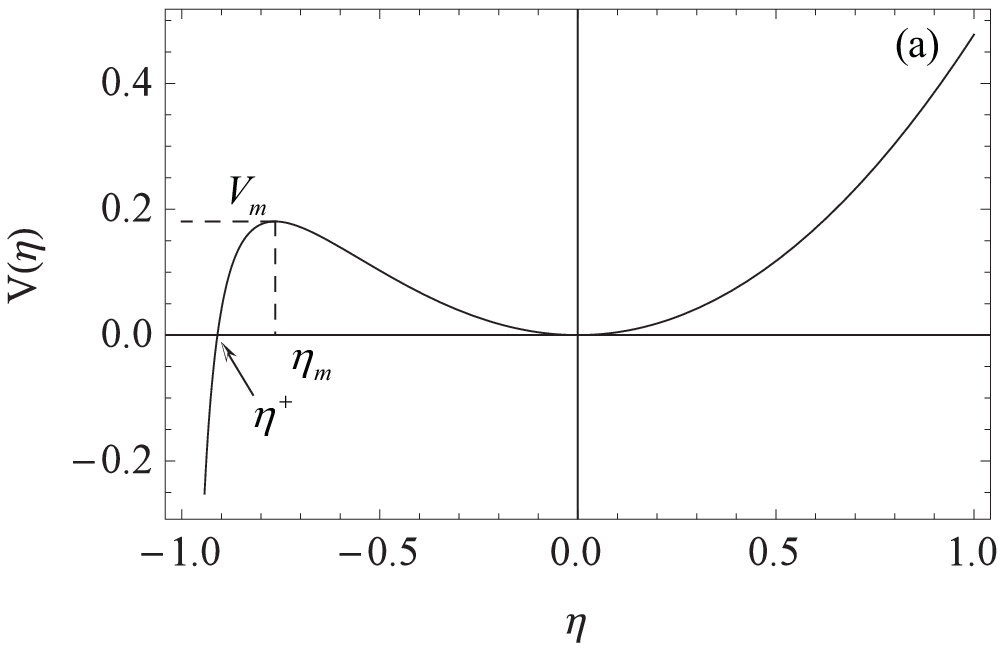} 
\par\end{centering}

\begin{centering}
\includegraphics[width=3.5in]{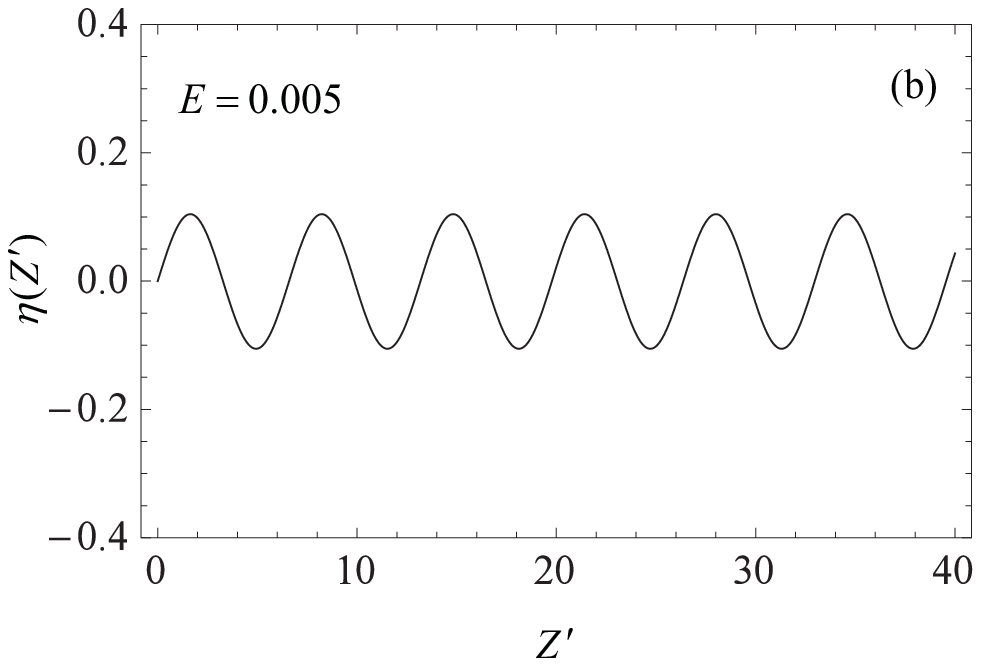}
\par\end{centering}

\protect\caption{Plots are shown for (a) $V(\eta)$ verses $\eta$, and (b) $\eta(Z^{\prime})$
verses $Z^{\prime}$, obtained from Eq.\,(94) for $M^{2}=0.09,$
$\eta'(0)=0.1,$ $E=1/2[\eta'(0)]^{2}=0.05,$ $\eta^{+}=-0.91,$ $\eta_{m}=-0.764$
and $V(\eta_{m})=0.181.$}

\label{Fig16} 
\end{figure}

\begin{figure}[ptb]
\begin{centering}
\includegraphics[width=3.5in]{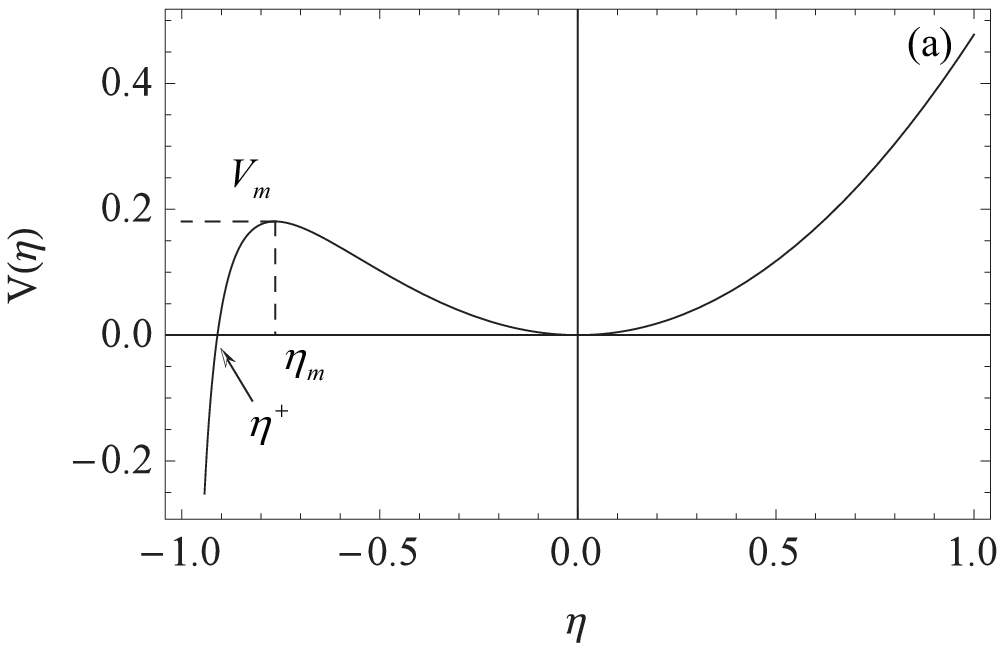} 
\par\end{centering}

\begin{centering}
\includegraphics[width=3.5in]{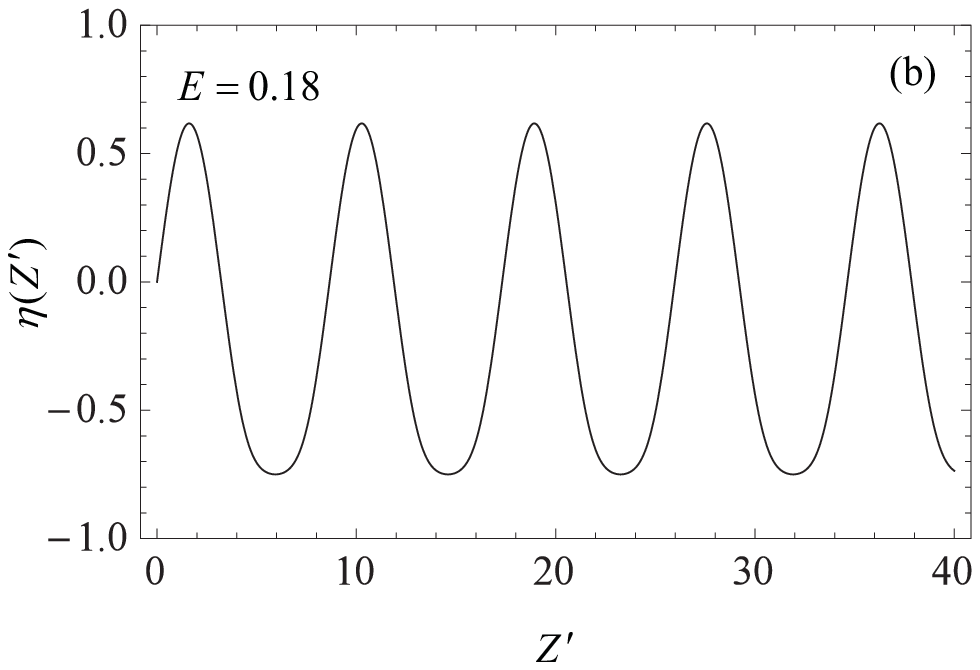}
\par\end{centering}

\protect\caption{Plots are shown for (a) $V(\eta)$ verses $\eta$, and (b) $\eta(Z^{\prime})$
verses $Z^{\prime}$, obtained from Eq.\,(94) for $M^{2}=0.09,$
$\eta'(0)=0.6,$ $E=1/2[\eta'(0)]^{2}=0.18,$ $\eta^{+}=-0.91,$ $\eta_{m}=-0.764$
and $V(\eta_{m})=0.181.$}

\label{Fig17} 
\end{figure}

Recall that the primed variables $\left(Z',V_{z}',T'\right)$ are
related to $\left(Z,V_{z},T\right)$ by $Z'=Z-MT$, $V_{z}'=V_{z}-M$
and $T'=T$, where $M=const.$ is the dimensionless velocity of the
traveling wave relative to to the unprimed frame. Therefore, for a
nonlinear wave that is time stationary ($\partial/\partial T'=0$)
in the primed variables, it is reasonable to identify $W_{U}'$ with
$W_{U}'=\left(1/2\right)M^{2}$ for a monoenergetic beam. In this
case, we make the identification $2W_{U}'<1$, so the condition for
Eqs.\,(\ref{eq:94}) and (\ref{eq:95}) to have nonlinear periodic
solutions for $\eta\left(Z'\right)$ {[}see Eq.\,(\ref{eq:98}){]}
can be expressed as

\begin{equation}
M^{2}<1\,.\label{eq:100}
\end{equation}
Typical numerical solutions for $\eta\left(Z'\right),$ obtained by
integrating Eq.\,(\ref{eq:94}) with $V\left(\eta\right)$ specified
in Eq.\,(\ref{eq:97}), are illustrated in Figs.\,\ref{Fig14}-\ref{Fig17}
for several choices of $M^{2}<1$ and different values of effective
energy level $E$. These correspond to: $M^{2}=0.5$, $E=0.005$,
$\eta_{m}=-0.360$, and $V\left(\eta_{m}\right)=0.014$ (Fig.\,\ref{Fig14});
$M^{2}=0.5$, $E=0.054883$, $\eta_{m}=-0.360$, and $V\left(\eta_{m}\right)=0.014$
(Fig.\,\ref{Fig15}); $M^{2}=0.09$, $E=0.05$, $\eta_{m}=-0.764$,
and $V\left(\eta_{m}\right)=0.181$ (Fig.\,\ref{Fig16}); and $M^{2}=0.09$,
$E=0.18$, $\eta_{m}=-0.764$, and $V\left(\eta_{m}\right)=0.181$
(Fig.\,\ref{Fig17}). Figures \ref{Fig14}-\ref{Fig17} illustrate
several interesting trends in the nonlinear periodic wave solutions
for $\eta\left(Z'\right)$. {[}These should be compared with the nonlinear
periodic wave solutions in Figs.\,\ref{Fig2}-\ref{Fig5} obtained
in Sec.\,\ref{sec:Waterbag} for the kinetic waterbag model.{]} First,
for smaller values of $M^{2}$, the potential wells are deeper and
broader (compare Figs.\,\ref{Fig14}a and \ref{Fig15}a with Figs.\,\ref{Fig16}a
and \ref{Fig17}a). Furthermore, the nonlinear wave amplitudes tend
to be large for sufficiently large values of energy level $E$ in
the potential well (compare Figs.\,\ref{Fig15} and \ref{Fig17}
with Figs.\,\ref{Fig14} and \ref{Fig16}).

\section{Conclusions \label{sec:Conclusions}}

In this paper, the 1D kinetic model developed in Ref.\,\cite{14davidson2004self}
was used to describe the nonlinear longitudinal dynamics of intense
beam propagation, allowing for moderate-to large-amplitude modulation
in the charge density of the beam particles. Particular emphasis has
been placed on investigating detailed properties of nonlinear pulse-like
(soliton) and periodic traveling-wave disturbances propagating with
constant normalized velocity $M=const.$ relative to the beam frame.
The 1D kinetic formalism \cite{14davidson2004self} was briefly summarized
in Sec.\,\ref{sub:Theoretical-Model-and}, and exact (local and nonlocal)
nonlinear conservation constraints were derived in Sec.\,\ref{sub:Conservation-Relations}
for the conserved particle number, momentum, and energy per unit length
of the beam, making use of the nonlinear Vlasov equation for $F_{b}\left(z,p_{z},t\right)$
in Eq.\,(\ref{eq:1}) and the expression for $\left\langle E_{z}\right\rangle \left(z,t\right)$
in Eq.\,(\ref{eq:2}). Removing the assumption of weak nonlinearity
made in Ref.\,\cite{15davidson2004korteweg}, Sec.\,\ref{sec:Waterbag}
made use of the fully nonlinear kinetic waterbag model to investigate
detailed properties of traveling nonlinear disturbances propagation
with velocity $M=const.$ relative to the beam frame. In normalized
variables, $Z'=Z-MT$ and $T'=T$, the waveform of the disturbance
was assumed to be time-stationary ($\partial/\partial T'=0$) in the
frame moving with velocity $M=const.$ Nonlinear solutions were examined
over a wide range of system parameters for both traveling-pulse (soliton)
and nonlinear traveling wave solutions in which the modulation in
beam density was large-amplitude, corresponding to a strongly bunched
beam. Finally, in Sec.\,\ref{sec:Kinetic} we examined the kinetic
model based on Eqs.\,(\ref{eq:9}) and (\ref{eq:10}) {[}equivalent
to Eqs.\,(\ref{eq:1}) and (\ref{eq:2}){]} for an even broader class
of distribution functions $F_{b}\left(z,p_{z},t\right)$. The analysis
in Sec.\,\ref{sec:Kinetic} parallels the original Bernstein-Greene-Kruskal
(BGK) formulation of BGK solutions to the 1D Vlasov-Poisson equations
\cite{25bernstein1957exact,26SeeMethod2}, except for the fact that
Eq.\,(\ref{eq:10}), which connects the effective potential $\left\langle \phi\right\rangle \left(z,t\right)$
to the line density $\lambda_{b}\left(z,t\right)$, has a very different
structure than the 1D Poisson's equation used in the original BGK
analysis. Depending on the choices of trapped-particle and untrapped-particle
distribution functions, the kinetic model described in Sec.\,\ref{sec:Kinetic}
supports a broad range of nonlinear pulse-like (soliton) solutions
and periodic traveling-wave solutions that have stationary waveform
in a frame of reference moving with velocity $M=const.$ relative
to the beam frame. Similar to Sec.\,\ref{sec:Waterbag}, the modulation
in beam line density can have large amplitude, corresponding to a
strong bunching of the beam particles. Specific examples were considered
in Sec.\,\ref{sec:Kinetic} corresponding to nonlinear periodic traveling
wave solutions of Eqs.\,(\ref{eq:9}) and (\ref{eq:10}). 
\begin{acknowledgments}
This research was supported under the auspices of U.S. Department
of Energy Contract No. DE-AC02-09CH11466 with the Princeton Plasma
Physics Laboratory.
\end{acknowledgments}


\clearpage{}


\end{document}